\documentclass[acmsmall, screen]{acmart}

\usepackage{tabularx}
\usepackage{booktabs}
\usepackage{array}
\usepackage{hyperref}
\usepackage{enumitem}
\usepackage{soul}
\usepackage{tikz}
\usepackage{graphicx}
\usepackage{longtable}
\usepackage{makecell}
\usepackage{multirow}

\newif\ifshowchanges
\showchangestrue
\showchangesfalse  

\ifshowchanges
    \usepackage[commandnameprefix=ifneeded]{changes}
    \setdeletedmarkup{\textcolor{gray}{\sout{#1}}}
\else 
    \newcommand{\added}[1]{#1}
    \newcommand{\deleted}[1]{}
    
\fi

\AtBeginDocument{%
  \providecommand\BibTeX{{%
    \normalfont B\kern-0.5em{\scshape i\kern-0.25em b}\kern-0.8em\TeX}}}

\setcopyright{acmcopyright}
\copyrightyear{2018}
\acmYear{2018}
\acmDOI{XXXXXXX.XXXXXXX}

\acmConference[Conference acronym 'XX]{Make sure to enter the correct
  conference title from your rights confirmation emai}{June 03--05,
  2018}{Woodstock, NY}
\acmISBN{978-1-4503-XXXX-X/18/06}

\begin{document}
\title{Rhetorical XAI: Explaining AI's Benefits as well as its Use via Rhetorical Design} 

\author{Houjiang Liu}
\affiliation{%
  \institution{School of Information, University of Texas at Austin}
  \city{Austin}
  \state{Texas}
  \country{USA}
}

\author{Yiheng Su}
\affiliation{%
  \institution{School of Information, University of Texas at Austin}
  \city{Austin}
  \state{Texas}
  \country{USA}
}

\author{Matthew Lease}
\affiliation{%
  \institution{School of Information, University of Texas at Austin}
  \city{Austin}
  \state{Texas}
  \country{USA}
}




\begin{abstract}
We explore potential benefits of incorporating Rhetorical Design into the design of Explainable Artificial Intelligence (XAI) systems. 
\deleted{While XAI is traditionally framed around explaining individual predictions or overall system behavior, explanations also function as a form of argumentation, shaping how users evaluate system perceived usefulness, credibility, and how they develop appropriate levels of trust and adoption.}
\added{While XAI is traditionally framed around explaining individual predictions or overall system behavior, explanations may also function as rhetorical arguments that shape how users evaluate a system’s usefulness and credibility, and how they develop appropriate trust for adoption.}
\added{In real-world, in-situ interactions, explanations can thus produce experiential and affective rhetorical effects that are not fully captured by traditional XAI design goals that focus primarily on how AI works.}
\added{To address this gap, we propose \textit{Rhetorical XAI}, which bridges two explanatory goals: \textit{how AI works} and \textit{why AI merits use}.}
\added{Rhetorical XAI comprises three appeals in explanation design: \textit{logos}, which aligns technical logic with human reasoning through visual and textual abstractions; \textit{ethos}, which establishes contextual credibility based on the explanation source and its appropriateness to the decision task; and}
\added{\textit{pathos}, which engages user emotionally by framing explanations around their motivations, expectations, or situated needs during interaction.}
\added{We conduct a narrative review synthesizing design strategies from prior XAI work aligned with these three rhetorical appeals, highlighting both opportunities and challenges of integrating rhetorical design into XAI.}

\deleted{Rhetorical Design offers an analytical framework to analyze the communicative role of explanations between AI systems, and end-users , focusing on : (1) logical reasoning conveyed through different types of explanations, (2) credibility projected by the system and its developers, and (3) emotional resonance elicited in users.} 
\deleted{Together, these rhetorical appeals helps us understand}
\deleted{how explanations influence people perceptions and facilitate AI adoption across and within different collaborative and social contexts.}

\end{abstract}



\maketitle

\section{Introduction} \label{sec:Introduction}


Modern AI systems are notoriously opaque, limiting efforts to understand or audit their behaviors \citep{Cambria2023-ld, Saeed2023-vx}. In response, Explainable Artificial Intelligence (XAI) aims to foster trust and accountability in AI systems by making their decision-making processes more transparent \citep{Busuioc2021-gp, Jacovi2021-id}. XAI encompasses techniques to support human understanding of both local predictions (individual decisions) and global behaviors \citep{Mohseni2021-bw, Cambria2023-af, Cabitza2023-hh, Lai2023-bu}. 


\added{However, XAI is not solely a technical challenge of producing faithful rationales of model behavior. 
XAI is also a communication problem because explanations are situated messages whose interpretation is mediated by who presents them, how they are framed, and who must act on them. 
In practice, different stakeholders leverage explanations to pursue distinct goals \citep{Kim2024-hp, Hoffman2021-ek, Preece2018-ql, Tomsett2018-ij, Mohseni2021-bw}. Developers primarily engage with explanations authored for technical audiences to debug model errors and biases \citep{Hohman2018-jf, Molnar2020-uo}. 
Ethicists and policymakers rely on explanations framed to justify system behavior when assessing AI safety and accountability \citep{Busuioc2021-gp}. 
End-users draw on explanations to decide how to appropriately leverage AI to improve task performance and efficiency \citep{Zhang2020-pf, Lai2023-bu, Vasconcelos2023-cm}.
Together, these cases illustrate that explanation effectiveness is not intrinsic, but depends on the interplay between the explanation’s source, its framing, and the recipient’s role and task context.}


\deleted{HCI researchers investigate how explanations can foster user appropriate trust \mbox{\citep{Zhang2020-pf, Lai2023-bu, Vasconcelos2023-cm}} and support complementary human-AI teaming \mbox{\citep{Bansal2020-zw}}. Stakeholders with diverse objectives tend to evaluate explanations through different lenses, shaped by their specific roles and needs. In particular, end-users tend to judge explanations alongside more pragmatic and affective dimensions such as utility, trustworthiness, and alignment with personal values \mbox{\citep{Vera_Liao2022-ro}}. Therefore, a faithful and logically accurate explanation that neglects human-centered considerations may be completely disregarded.} 


In this article, we \deleted{argue that designing explanations for effective adoption} expand the conceptual scope of XAI beyond explaining \textit{how AI works} towards also articulating \textit{why AI merits use}.
An AI system is just one of many possible solutions to a user problem, 
\added{so AI adoption warrants justification.}
\added{Through this lens,} AI explanations can serve a \textbf{rhetorical} function by communicating why an AI system is beneficial, credible, and appropriate in a given context.
\deleted{Applications can help support critical evaluations of AI decisions, assess system credibility, and calibrate appropriate user trust.}
\added{From this perspective, explanations are not technical specifications but designed artifacts. They produce diverse rhetorical effects (e.g., experiential, affective, and even irrational forms of persuasion) that are not fully captured by traditional XAI design goals focused primarily on explaining how AI works.}

\added{
To expand the dominant technical understanding of XAI and situate it within a social perspective, we propose \textbf{Rhetorical XAI}, an analytical framework rooted in \textit{Rhetorical Design} that extends XAI beyond explaining how AI works to also account for why AI systems merit use. Rhetorical XAI characterizes explanation design through three rhetorical appeals:
}
(1) \textbf{logos}, \added{which aligns technical logic with human reasoning through visual and textual abstractions}\deleted{how different explanation types appeal to distinct form of human reasoning}; 
(2) \textbf{ethos}, \added{which establishes contextual credibility based on the explanation source and its appropriateness to the decision task;} \deleted{how the AI system and its developers convey trustworthiness}
and (3) \textbf{pathos}, \added{which engages user emotionally by framing explanations around their motivations, expectations, or situated needs during interaction.} \deleted{how different affective reactions elicited by explanations influence user perception and engagement with the system}\added{Using this framework, we synthesize prior XAI design strategies along three rhetorical dimensions and across the two explanatory goals. In doing so, we situate Rhetorical XAI within critical discourse on persuasive technologies, foregrounding questions of (in)appropriate explanation use in social and collaborative contexts.}

\added{
The remainder of this article is structured as follows. 
Section~\ref{sec:Background} situates XAI as a communicative problem by reviewing relevant communication theories in HCI and CSCW, motivating the inclusion of rhetorical perspectives for explanation design.
Section~\ref{sec:Narrative Review} outlines our narrative review methodology, detailing the procedures for literature selection, coding, and synthesis to support analytical transparency.
Section~\ref{sec:XAI design} reviews existing XAI design paradigms and reveals a prevailing focus on explanations as informative tools for understanding how AI works, rather than as rhetorically designed artifacts that justify why AI merits use. To address this limitation, Section~\ref{sec:Framework} introduces the Rhetorical XAI framework (Table~\ref{tab:rhetorical AI explanations}), which characterizes explanation design along three rhetorical dimensions: \textit{logical reasoning} (logos), \textit{credibility} (ethos), and \textit{emotional resonance} (pathos). 
Section~\ref{sec:Results} presents our synthesis of prior XAI design strategies across different rhetorical appeals and explanatory goals (Table~\ref{tab:rhetorical_strategies}). 
Finally, Section~\ref{sec:Discussion} critically examines the benefits and risks of identified rhetorical strategies, outlining implications and open challenges for responsible AI adoption in social and collaborative contexts.
}


\deleted{This article is structured as follows. Section \ref{sec:Background} reviews key surveys core technical concepts and existing design paradigms in XAI, establishing the basis for underscoring the overlooked potential of incorporating rhetorical design into the current XAI landscape. In Section \ref{sec:Rhetoric}, we introduce and adapt rhetorical design into XAI through a conceptual framework. In Section \ref{sec:Method}, we describe how we apply our rhetorical framework to guide the qualitative coding process when reviewing prior XAI designs. We present our coding results in Section \ref{sec:Results}, elaborating on the identified design strategies. In Section \ref{sec:Discussion}, we critically examine the benefits and limitations of these strategies and offer recommendations for future XAI research. Finally, in Section \ref{sec:Conclusion}, we summarize our contributions toward advancing rhetorical XAI.}

\section{Theoretical Background} \label{sec:Background}
\added{
XAI is not solely a technical challenge on how to faithfully explain model behaviors, but also a fundamentally communicative problem in which explanations function as designed artifacts that shape how AI systems are understood and acted upon. Section~\ref{subsec:communication} reviews how HCI researchers have drawn on communication theories to examine how explanations are produced, framed, and interpreted across different stakeholders and contexts. Section~\ref{subsec:rhetorical HCI} then draws on rhetoric, a communication tradition less examined in XAI, to position explanations as persuasive artifacts that justify not only how AI works, but why AI merits use.
}

\subsection{XAI as a Communication Problem} \label{subsec:communication}

\added{
A foundational formulation of XAI as a communicative problem was offered by \citet{Miller2019-zh}, who situated XAI in the cognitive and social processes through which humans naturally constructed explanations.
Drawing on Peirce’s theory of abduction \citep{peirce1997pragmatism}, Miller argued that contrastive explanations (``\textit{why P rather than Q?}'') best support human causal inference because they reflected abductive reasoning, which favored the most plausible explanation among alternatives.
Building on the linguistic tradition of \textit{pragmatics}, Miller further incorporated Grice's maxims \citep{Grice1975-zz} and conversational models \citep{Walton2004-io, Hilton1990-fm, Antaki1992-jm}
to frame AI explanations as cooperative dialogues for shared understanding.
This perspective has substantially influenced subsequent XAI research, including technical methods for generating contrastive explanations \citep{Karimi2023-ef, Jacovi2021-me}, empirical evaluations of their efficacy \citep{Van_der_Waa2021-he, Bucinca2025-kb}, and designs of conversational explanatory systems \citep{explainDialogue2022, Gaole2025-wo}.}

\added{Although Miller framed these theoretical relations as the “social sciences” of XAI, his account primarily conceptualized explanation as a form of logical discourse that emphasized causal coherence and linguistic pragmatics.}
\added{In contrast, scholars in human–computer interaction (HCI) and computer-supported cooperative work (CSCW) have adopted broader communication perspectives that situated AI explanations within social, organizational, and collaborative contexts.
For example, drawing on \citet{Luhmann1992-eb}’s multi-layer cybernetics, \citet{Keenan2023-oc} argued that the meaning of AI explanations emerged from complex \textit{N}-order interpretations within social groups, rather than from simple dyadic human–AI dialogue.
\citet{Ehsan2021-id} built on \citet{Erickson2000-mg}'s social translucence to reveal organizational implications of AI explanations
grounded in sociocultural communication practices \citep{Heath2000-yh, Orlikowski1994-yv}.}

\added{Communication scholars complemented these perspectives by highlighting an imbalance in how XAI research attended to different elements of the communication process.
For example, \citet{Xu2024-mx} observed that most HCI work on XAI adopted a predominantly receiver-focused perspective, emphasizing how user needs, human factors, and organizational contexts affected explanation effectiveness. 
In contrast, they noted that explanation sources were commonly discussed only at the dataset level, leaving the roles of system developers, designers, and deploying organizations largely unaddressed.}

\added{
In summary, viewing XAI as a communicative problem positions explanations as designed artifacts whose interpretations arise from the interplay between explanation sources, explanatory formats, and target recipients within social contexts \citep{Miller2019-zh, Ehsan2021-id, Xu2024-mx}.
}

\subsection{The Communication Root in HCI, CSCW, and Their Rhetorical Standpoints} \label{subsec:rhetorical HCI}

\added{
Building on the view of XAI as a communicative problem, this section situates rhetoric within the broader communication traditions that have shaped HCI and CSCW. 
While prior work has drawn on semiotic, pragmatic, and sociocultural theories to examine how meaning is interactively encoded and negotiated, rhetoric foregrounds the deliberate design of communicative forms to persuade, justify, and establish credibility.
We draw on this rhetorical lineage in HCI to position AI explanations as designed artifacts whose form, framing, and presentation influence not only how systems are understood, but also how they are evaluated and adopted in practice.
}

\subsubsection{The Semiotic and Pragmatic Focus in HCI} \label{subsubsec:pragmatic}
\added{
\citet{Barbosa2024-lb} traced the historical role of language and communication in HCI primarily through the lenses of \textit{semiotics} and \textit{pragmatics}. 
Semiotics attends to how meaning is encoded in static representational elements such as buttons, text, and images, while pragmatics extends this focus to how meaning is constructed through users’ actions and interaction.
For example, Grice’s maxims \citep{Grice1975-zz} offer a pragmatic framework based on quantity, quality, relation, and manner for designing and evaluating conversational behavior.
The Speech Act theory \citep{Winograd1987, Searle1969-om} builds on this pragmatic view by treating language as performative, enabling utterances to function as actions that trigger system behavior rather than merely convey descriptive information.
\citet{Dubberly2009-gm} drew on \textit{cybernetics} to conceptualize interaction as an ongoing conversational system, distinguishing between reactive, self-regulating, and learning behaviors that emerge through feedback over time. 
} 

\subsubsection{The Sociocultural and Critical Focus in CSCW} \label{subsubsec:sociocultural} 

\added{
CSCW researchers focus on the technological challenges that arise in collaborative work environments, particularly those involving interdependent activities, temporal coordination, and organizational structures \citep{Star1989-rp, Ackerman2000-sv}.
In these settings, pragmatic theories might not fully account for the situated and evolving nature of real-world collaborative work. 
For example, \citet{Bowers1988-gc} shown that while classic Speech Act theory treated interaction as a sequence of discrete, well-defined utterances, real-world coordination was organized around shared activity contexts and situational cues that rendered turn-taking fluid and often implicit.
Similarly, \citet{Suchman1993-bd} argued that categorizing interaction into predefined speech utterances (e.g., requests, commitments, promises) imposed rigid control structures that failed to accommodate the contingent and adaptive nature of work.}

\added{In response to these socio-technical gaps \citep{Ackerman2000-sv}, CSCW scholars have increasingly drawn on \textit{sociocultural} and \textit{critical} communication theories to better account for how meaning is encoded and negotiated in practice.
Examples include \citet{Orlikowski1994-yv}’s concept of technological frames, which built on organizational sensemaking theory \citep{Weick1990-hn} to examine how different social groups interpreted technologies, as well as \citet{Biocca2003-fx}’s application of social presence theory to understand user satisfaction in online media contexts.}

\subsubsection{Rhetorical Theories from Language to HCI} \label{subsubsec:rhetorical theories}

\added{Building on the above communication paradigms, we turn to rhetoric as another HCI tradition that shifts attention from interpretation and coordination to persuasion and justification.} 

\added{Rhetoric foregrounds persuasion. As \citet{Craig1999-vj} explained, among seven communication traditions, Semiotics examined ``intersubjective mediation by signs'' and Phenomenology focused on ``dialogue experience[s]'' alongside four additional traditions, including Cybernetic, Sociopsychological, Sociocultural, and Critical. In contrast, Rhetoric centered on ``how the artful use of discourse [serves] to persuade audiences.''} Aristotle first defined rhetoric as ``the faculty of observing in any given case the available means of persuasion'' \citep{Aristotle1926-uq}. He stated that an argument persuades not only through logical reasoning (\textit{logos}) but also through the speaker’s credibility (\textit{ethos}) and the emotional influence exerted on the audience (\textit{pathos}). For example, a genuine diamond ring (strong \textit{logos}) might be perceived as fake if presented by someone in severe financial hardship (weak \textit{ethos}). Similarly, straightforward truths (strong \textit{logos}) might fail to convince a child who is crying and having a tantrum (weak \textit{pathos}). 

\added{
Traditionally focused on discourse, rhetoric has been adopted in related domains such as advertising \citep{Burke1969-rd}, narrative fiction \citep{Booth1983-ek}, social media \citep{Carnegie2009-qr}, and more recently in HCI.
Here, \citet{Boyarski1994-je} framed interaction design as persuasive discourse, where design choices shape system understanding and use through logical structure, implied voice, and affective elements.
}
\added{
Similarly, \citet{Buchanan1985-qy} viewed electronic products as rhetorical artifacts designed to attract users through technological reasoning (\textit{logos}), manufacturer credibility (\textit{ethos}), and emotional resonance (\textit{pathos}).
Beyond artifacts, \citet{Brummett1999-ao} conceptualized machines as rhetorical social agents that participated in meaning-making through functional utility (\textit{logos}), aesthetic and material qualities (\textit{ethos}), and cultural resonance (\textit{pathos}).
}
\added{
Notably, rhetoric challenges assumptions of idealized rationality that often underpin pragmatic theories (Section \ref{subsubsec:pragmatic}). 
As \citet{Barbosa2024-lb} noted, Grice's maxims \citep{Grice1975-zz} assumed ``people engaged in communicative interaction will do their best to get their message across, and in doing so will abide by a number of conversational conventions, or \textit{maxims}.'' 
However, in practice, communication often departs from these ideals.
While rhetoric incorporates rational argumentation as one persuasive strategy (e.g., \textit{logos}), it also highlights additional appeals such as \textit{ethos} and \textit{pathos}, which account for affective, situational, and non-rational influences inherent in social interactions.
Accordingly, rhetoric offers a valuable lens for scrutinizing HCI designs in practical settings where user behavior cannot be assumed to follow idealized models of rationality.
}

\subsubsection{Rhetorical Design}
\added{
HCI practitioners have developed concrete design strategies to enact different rhetorical appeals. 
In web interface design, for example, emotional engagement (\textit{pathos}) is facilitated by visually prominent elements such as strong contrast, surprising details, and aesthetic signals. Credibility (\textit{ethos}) is established through reassurances and forms of social proof, while logical appeal (\textit{logos}) is conveyed through clear and verifiable information \citep{Ibrahim2013-vo, Sosa-Tzec2014-cs, Sosa-Tzec2015-ke}. 
Similarly, information visualization employs linguistic devices (e.g., irony, analogy) alongside visual techniques (e.g., metaphor, contrast, categorization) to influence perceptions of trustworthiness, emotional resonance, and logical coherence \citep{Hullman2011-sb}.
Related work on infographics also leverages rhetorical strategies to communicate complex social and medical topics, including recycling and sustainability \citep{Hua2021-zs}, public health statistics \citep{Meuschke2022-rw}, and civic participation \citep{Dork2013-ir, Claes2017-pp}. 
On the other hand, HCI researchers have examined how rhetorical techniques can be applied in harmful ways, resulting in coercion or ``dark patterns'' that actively deceive users for the benefit of other parties. 
For instance, \citet{Gray2018-bt} illustrated how interface designs can deliberately exploit user cognitive and emotional vulnerabilities to steer them toward unintended decisions, a concern that has been extended to XAI by \citet{chromik2019dark}.
}

\subsubsection{Advocate for Rhetorical XAI}
\added{
Building on the view of XAI as a communicative problem, this article adopts a rhetorical perspective that conceptualizes AI explanations as purposively designed artifacts rather than neutral descriptors of model behavior. 
This rhetorical lens thus extends existing HCI rhetoric traditions into XAI by shifting attention from explaining how AI systems work to also examining how explanations persuade users that AI systems merit use.
By doing so, this rhetorical perspective makes explicit the experiential, affective, and non-ideal influences on users that are often overlooked by need-driven XAI design goals centered on technical understanding.
} 

\added{
Many existing XAI studies already report, either explicitly or implicitly, appeals that can be understood through a rhetorical lens, including work on persuasive AI advice adoption \citep{Dragoni2020-tn}, experiential explanation design \citep{El-Zanfaly2023-da, Bryan-Kinns2024-kk}, tangible and sensory explanation experiences \citep{Ghajargar2023-zl, Alvarado2022-ys}, as well as dark patterns associated with explanation interfaces \citep{chromik2019dark}. 
However, these contributions remain fragmented across domains and lack a unifying account that theoretically characterizes design strategies alongside different rhetorical dimensions and explanatory goals. 
To address this gap, this article introduces a rhetorical framework (Section~\ref{sec:Framework}) that synthesizes existing explanation design strategies from prior XAI research (Section~\ref{sec:Results}).
}


\noindent\deleted{In this section, we briefly review key technical concepts (Section \ref{subsec:XAI terminologies}) and design paradigms (Section \ref{subsec:user-centered XAI}, \ref{subsec:stakeholder-centered XAI}, and \ref{subsec:sociotechnical XAI}) in XAI. This summary draws primarily from existing survey papers, position papers, and design frameworks in XAI, serving as background to contextualize our perspective on rhetorical design.}

\deleted{Both AI and HCI researchers have acknowledged that making AI systems interpretable to people can foster more trustworthy and accountable use of AI \mbox{\citep{Markus2021-fz, Molnar2020-uo, Jacovi2021-id, Busuioc2021-gp}}. \mbox{\citet{Doran2017-gd}} categorize AI systems into three types --- \textit{opaque}, \textit{interpretable}, and \textit{comprehensible} --- based on the level of transparency they offer in explaining automated decisions. In this taxonomy, interpretability refers to how well users can understand the internal processes by which an AI system transforms inputs into outputs. Other taxonomies of interpretability frame this input-output relationship through alternative lenses such as explanation scope, algorithmic techniques, or interactive format.} 

\deleted{A common taxonomy based on the explanation scope is whether AI explanations are \textit{local} or \textit{global} \mbox{\citep{Molnar2020-uo, Mohseni2021-bw, Dwivedi2023-oc}}. \textit{Local} explanations explain why a specific input maps to an observed output. In contrast, \textit{global} explanations provide an overarching view of how the AI generally maps inputs to outputs. Orthogonally, scope can also vary by model generality. Another foundational taxonomy differentiates \textit{model-specific} techniques, which are tailored to a particular model, from \textit{model-agnostic} techniques, which can be applied to any model.} 

\deleted{Different XAI techniques lead to diverse types of explanations, which HCI researchers have evaluated with respect to their effects on users. These explanation types include \textit{example-based}, \textit{feature-based}, \textit{counterfactual}, \textit{rule-based}, and \textit{concept-based} explanations (see \textbf{Appendix \ref{appendix:types of AI explanations}} for further details). Each supports a distinct form of logical reasoning, reflecting the underlying algorithmic mechanisms that generate the explanation \citep{Wang2019-we}.} 

\deleted{Given the wide range of human-AI interactions enabled by explanations \mbox{\citep{Bertrand2023-xy, Xu2023-qd, Chromik2021-ef}}, many user studies adopt two primary forms of explanations --- \textit{static} vs. \textit{interactive} --- to examine how different explanation types influence user acceptance or foster appropriate trust in individual AI predictions. Static explanations typically present fixed outputs for given inputs and allow participants to engage with them over a limited number of trials. In contrast, interactive explanations enable users to actively engage with the system by adjusting inputs or model parameters, allowing them to explore alternative scenarios and gain a deeper understanding of the model’s behavior \mbox{\citep{Liu2021-ds, Bertrand2023-xy, Chromik2021-ef, Nguyen2018-az}}.}

\deleted{While there is debate about the necessity of AI explanations, many researchers view XAI as a promising pathway for promoting safe, trustworthy, and accountable use of AI. For instance, when humans collaborate with AI to perform different tasks, explanations can help users understand model behavior, enabling them to judge when to accept or reject AI recommendations. This, in turn, fosters appropriate user trust and supports complementary human-AI teaming \mbox{\citep{Lee2004-kt, Yang2020-wm, Wang2021-hv, Schemmer2023-zy}}. To support cooperation, HCI researchers typically draw on human-centered principles when designing XAI systems (HCXAI). HCXAI emphasizes identifying user needs and contexts of use to inform whether, when, and how different explanations should be presented \mbox{\citep{Ehsan2020-gz, Ehsan2023-HCXAI, Ehsan2024-HCXAI}.}} 

\deleted{Building on this perspective, the following sections briefly introduce XAI design paradigms including user-centered (Section \ref{subsec:user-centered XAI}), stakeholder-centered (Section \ref{subsec:stakeholder-centered XAI}), and sociotechnical approaches to explanation (Section \ref{subsec:sociotechnical XAI}).}

\deleted{HCI scholars studying XAI have focused on how different types of explanations can support users in understanding AI systems more effectively \mbox{\citep{Hagras2018-hx, Liao2020-da, Vera_Liao2021-zi}}. \mbox{\citet{Hagras2018-hx}} first argued that XAI aimed to \textit{make AI understandable to humans}. One strategy toward this goal is to reframe technical descriptions of AI decision-making into user-friendly language. For example, consider explaining  how an autonomous car changes speeds. Instead of a technical explanation such as ``if the distance to the car ahead is less than 2.5m and the road is 10\% slippery then reduce car speed by 25\%,'' a more human-like natural language rephrase might be ``if the distance to the car ahead is low and the road is slightly slippery, then slow down.'' In general, researchers have developed various types of explanations (see Appendix \ref{appendix:types of AI explanations}) that employ visualizations and natural language to make AI more accessible and understandable to people \mbox{\citep{Lai2023-bu, Robbemond2022-px}}.} 

\deleted{Given the diversity of explanations, researchers have proposed different frameworks to help designers select the most appropriate explanation type for a given context. For example, \mbox{\citet{Liao2020-da}} proposed an XAI ``question bank'' comprised of common user inquiries such as \textit{How} AI works, \textit{Why} AI made a specific prediction, \textit{Why not} an alternative prediction, \textit{What} AI would predict if the input changes, and \textit{How} to change the input to alter the AI prediction. Other researchers have adopted this question-informed framework to select AI explanations when designing XAI systems \mbox{\citep{Xin_He2023-dj, Ehsan2022-fz, Wang2023-pk}}. Other frameworks aim to bridge human understanding and machine explanations from a cognitive perspective, such as \mbox{\citet{Wang2019-we}} proposing to design AI explanations align with human cognitive processes, \mbox{\citet{Chen2022-cq}} incorporating task-specific human intuitions when assessing user understanding of explanations, and more \mbox{\citep{Zhang2019-wj, subhash2022makes, Chen2023-me}}.}

\deleted{Since AI systems serve diverse end-user groups, explanations can be tailored to their specific needs and contexts \mbox{\citep{Ehsan2024-lu, Kim2024-hp}}. 
For example, loan seekers and lenders have different explanatory needs for understanding how a credit scoring AI system functions \mbox{\citep{Bove2022-pd}}. Loan seekers might be particularly interested in why their financial record led to a rejection and how to alter the decision. Lenders might seek a global understanding of AI behavior across historical loan records to support financial analysis, enabling them to make more informed decisions about issuing loans to future applicants \mbox{\citep{Dikmen2022-hd}}. 
Similar distinctions are also prevalent in the medical domain between patients and healthcare providers \mbox{\citep{Burgess2023-dm, Kim2024-hp}}. Collectively, these cases highlight how domain expertise shapes explanatory needs \mbox{\citep{Mohseni2021-bw, Kim2024-hp}}. Lay users often prefer personalized and actionable guidance, while experts commonly seek system-level understanding.}

\deleted{While end-users are a central focus in XAI, they represent just one of the many stakeholders involved in developing, deploying, and adopting AI systems. To situate AI within this broader socio-technical ecosystem, explanations should also cater to stakeholders' distinct goals, responsibilities, and information needs. Two primary approaches for categorizing stakeholders are based on their \textit{roles} within the AI lifecycle \mbox{\citep{Tomsett2018-ij, Preece2018-ql, Hoffman2021-ek, Dhanorkar2021-ah}} and their level of \textit{AI knowledge} \mbox{\citep{Mohseni2021-bw}}.}

\deleted{In role-based stakeholder categorization, \mbox{\citet{Hoffman2021-ek}} demonstrated distinct explanatory needs between developers, policymakers, and domain users. They reported that developers mainly employed explanations to debug AI errors and identify potential algorithmic biases to improve overall predictive performance. Policymakers leveraged explanations to audit AI behavior in expected deployment settings and unexpected edge cases. Domain users used explanations to understand or counterfactually remediate unfavorable AI decisions. Other researchers also used role-based frameworks but with alternative groupings of stakeholders \mbox{\citep{Tomsett2018-ij, Preece2018-ql, Dhanorkar2021-ah}}.} 

\deleted{Complementary to stakeholder roles,\mbox{\citet{Mohseni2021-bw}} proposed a taxonomy of explanation needs based on varying levels of AI expertise. They distinguished between \textit{AI novices} (layman users) from \textit{AI experts} (engineers and developers), each leveraging explanations to suit different purposes. However, \mbox{\citet{Suresh2021-qq}} contended that both role-based and AI knowledge-based frameworks could be problematic, as overlapping explanation needs could challenge consistency between different stakeholder groups. Consequently, they proposed to organize stakeholder needs around \textit{Goals}, \textit{Objectives}, and \textit{Tasks} along the AI development cycle, aiming to offer a more cohesive and ontologically grounded vocabulary.} 

\deleted{While stakeholder frameworks might differ in structure and granularity, they all highlighted the need to contextualize AI explanations to specific goals and responsibilities of different stakeholder groups \mbox{\citep{Wolf2019-sv, Carroll2003-uw}}. Identified stakeholder needs then served as specifications to guide the technical development \citep{Kaplan2024-td} and interface design of XAI systems \mbox{\citep{Dhanorkar2021-ah, Ehsan2023-pu}}.}

\deleted{As researchers broaden their study focus from end-users to a wider range of stakeholders, this shift reflects a socio-technical perspective for studying XAI \mbox{\citep{Preece2018-ql, Ehsan2020-gz, Dhanorkar2021-ah}}. \mbox{\citet{Brandao2019-qi}} argued that AI experts often lack a natural understanding to connect the potential social meanings with models they create. \mbox{\citet{Ehsan2023-pu}} further illustrated this disconnection through a use case where an AI-powered sales prediction tool, despite its 90\% reported accuracy, was adopted by only 10\% of sellers due to either over-trust by less experienced users or disregard by experienced sellers. Similar cases have emerged in different areas, including customer-service relationship \mbox{\citep{Northey2022-it}}, journalist norms and values \mbox{\citep{Komatsu2020-xz}}, and frontline workers \mbox{\citep{Fox2023-gy}}.}

\deleted{The main issue described in the above use cases is that, current AI deployment did not account for organizational social needs \mbox{\citep{Ehsan2023-pu}}. In particular, users reported that verifying AI predictions often created added burdens for organizational work, such as having to learn new terminologies, consult technical experts, or manage concerns about reduced transparency. To address this, some XAI researchers argued that explanations should also account for users' roles, expertise, and collaborative practices within the organization \mbox{\citep{Ehsan2021-id, Ehsan2023-pu}}.}

\deleted{Thus far, Section~\ref{sec:Background} has outlined key technical concepts in XAI, covering commonly used terminologies and taxonomies (\ref{subsec:XAI terminologies}), as well as design paradigms of human-centered XAI, including user-centered (\ref{subsec:user-centered XAI}), stakeholder-centered (\ref{subsec:stakeholder-centered XAI}), and sociotechnical perspectives (\ref{subsec:sociotechnical XAI}). This section aims to differentiate our work by conceptualizing AI explanations as rhetorical artifacts that mediate communication between designers and end-users, beyond merely justifying outputs or revealing flaws.}

\deleted{While user-centered XAI (Section \ref{subsec:user-centered XAI}) prioritizes functional utility and cognitive human-AI alignment across different user objectives, we argue that pragmatic and affective responses equally shape understanding and adoption of AI \mbox{\citep{Dewey2005-cw, Buchanan1985-qy, Buchanan2001-ze}}. For example, when interacting with XAI systems, users frequently grapple with questions such as: Will this explanation help me achieve my goal better? Can I trust its rationale? Does it align with my ethical principles of using AI? These questions highlight that technical fidelity and cognitive alignment alone are insufficient; explanations must also resonate with user expectations, emotions, and values to be meaningful and persuasive.}

\deleted{More importantly, while the stakeholder-centered framework (Section \ref{subsec:stakeholder-centered XAI}) broadens the scope of XAI to include the goals and values of broader stakeholder groups, and socio-technical focus (Section \ref{subsec:sociotechnical XAI}) highlights the practical challenges of implementing XAI in an organizational setting, both still frame AI explanations primarily as tools either to justify AI results or to expose their flaws. We argue that this framing neglects the role of explanation as a communicative act in AI use, which is shaped by design and situated within a social context, inviting diverse interpretations around why AI appears logically sound, credibly useful, and emotionally appealing. When multiple stakeholders engage with explanations in collaborative settings, it is important to understand how different explanation designs shape their interpretations. These interpretations, in turn, influence perceptions of AI transparency, the negotiation of responsibilities, the development of trust, and the coordination of final AI adoption in practice.}

\deleted{Our reframing positions AI explanations not only as informative responses but also as \textbf{rhetorical artifacts} that mediate the relationship between AI systems and end-users. In other words, AI is just one of many possible solutions to a user problem. AI explanations thus must serve a persuasive and contextualizing role by communicating why an AI system is beneficial, credible, and appropriate in different contexts. From this perspective, a rhetorical XAI design can more effectively support stakeholders in evaluating automated decisions \mbox{\citep{Wang2022-lh, Bansal2020-zw, Nguyen2018-az}}, assessing system credibility \mbox{\citep{Binns2018-fs, Dodge2019-ss}}, and calibrating appropriate user trust \mbox{\citep{Markus2021-fz, Liao2022-jm}}.}

\deleted{In this article, we investigate the argumentative role of AI explanations in mediating user adoption of AI systems, and we synthesize how prior research has employed various design strategies to fulfill this rhetorical function. We draw on the concept of \textbf{Rhetorical Design}, a longstanding product and graphic design practice that examines how form, structure, and presentation influence user adoption. In the next section, we introduce the theoretical foundations of rhetorical design in HCI and describe how we adapt this lens to examine and inform explanation design.}


\section{Narrative Review Process} \label{sec:Narrative Review}
\added{
We conduct a narrative review to conceptualize our rhetorical framework and synthesize prior XAI design strategies. 
A narrative review offers an interpretive and critical account of prior research \citep{Ferrari2015-ay, Pare2017-mx, sukhera2022narrative}.
By emphasizing interpretive integration, narrative reviews are well-suited for theory-driven analysis across interdisciplinary fields, tracing how ideas evolve and reconciling heterogeneous rationales \citep{Ferrari2015-ay, Greenhalgh2018-ou}. 
}

\added{
A narrative review is not a systematic review, which prioritizes exhaustive coverage and methodological reproducibility through predefined protocols \citep{Moher1996-pr}. 
Scholarship across the natural and social sciences emphasizes that systematic and narrative reviews serve distinct but complementary purposes \citep{Baumeister1997-el, Stefanidi2023-gv}.
The emphasis of systematic reviews on exhaustive search and methodological reproducibility can be misaligned with the goals of narrative review, because data aggregation tends to flatten how ideas and theoretical positions are developed historically across studies \citep{Collins2005-qf, Moher1996-pr}.
Accordingly, \citet{Greenhalgh2018-ou} advocated for narrative reviews when authors prioritize interpretive and critical engagement with prior research over comprehensive enumeration.
\citet{Fok2025-ng} echoed this perspective through interviews with HCI scholars, who emphasized the long-standing value of narrative and semi-systematic reviews in enabling ``subjective, expertise-driven decisions and idiosyncratic strategies'' that ``motivate deeper structural and interpretative revisions.''
}

\added{
We chose to conduct a narrative review for two reasons. 
First, our contribution is not a comprehensive taxonomy of XAI design practices, but the development of a rhetorical framework and its use as an analytical lens to reinterpret explanation designs and their reported effects. This emphasis on theory-building and critical synthesis aligns well with a narrative approach.
Second, narrative reviews have a well-established precedent in HCI and CSCW for advancing conceptual and critical contributions. 
Examples include \citet{Das2023-fg}’s review of human-centered NLP fact-checking, \citet{Hsueh2024-vv}’s analysis of four epistemic positions in creativity-oriented HCI research, and \citet{Mohseni2021-bw}'s multidisciplinary survey on XAI design and evaluation. These studies similarly prioritize analytical insight over exhaustive coverage.
}

\added{
Following this tradition, our narrative review includes three main steps: (1) searching for and screening literature relevant to XAI design, (2) analyzing existing XAI design paradigms to inform the development of a rhetorical framework, and (3) applying this framework as a coding tool to identify concrete design strategies and corresponding rhetorical effects. 
This section details the first step by describing our literature search (Section \ref{subsec:search}) and screening procedures (Section \ref{subsec:screening}). To account for the interpretive nature of the narrative review process, we also include a reflexivity statement to acknowledge potential author bias (Section \ref{subsec:reflexivity}).
}

\subsection{Literature Search} \label{subsec:search}
\added{
Our literature search was conducted in two phases. The first phase, conducted in Summer 2024, involved keyword searches across four databases: Web of Science, ACM Digital Library, ScienceDirect, and Taylor \& Francis Online. We queried paper titles and abstracts using the keywords ``Explainable AI'' OR ``Interpretable AI'' AND ``Design''. The second phase, conducted in March–April 2025, focused on recent publications not captured in the initial search. Using the same search strategy, we primarily targeted XAI-relevant conferences on the ACM Digital Library, such as CSCW 2024, IUI 2025, and CHI 2025. We did not impose strict date limits in either search phase. We collected the corpus iteratively, reviewing search results until no more relevant papers were identified.}


\subsection{Literature Screening} \label{subsec:screening}

\added{
We organize the corpus into two groups: (I) survey and framework papers on designing XAI systems, and (II) evaluation papers that provide empirical evidence on the rhetorical effects of AI explanations. 
For group I, we searched the collected corpus for papers containing keywords ``frameworks'' OR ``principles''. These papers were coded to synthesize mainstream XAI design paradigms (Section \ref{sec:XAI design}). 
Next, empirical papers in group II were coded to identify rhetorical strategies and user effects (Section \ref{sec:Results}).
For transparency, we present an overview of the two-step literature screening in this section and provide additional statistics, tables, and detailed descriptions of each stage in Appendix \ref{appendix:review}.
}

\subsubsection{Phase 1: Screening Conceptual and Technical Literature}
\added{
Group I comprises surveys and framework papers from both technical and design perspectives, though our coding primarily focuses on the design-oriented contributions.
From this perspective, surveys either scope XAI design practices \citep{Bertrand2023-xy, Kaplan2024-td, Mohseni2021-bw} or propose a new framework for future XAI design \citep{Wang2019-we, Miller2023-nc, Mohseni2021-bw}. 
Empirical framework papers include some that categorize high-level design guidance from interviews \citep{Liao2020-da, Brennen2020-kw} and proposed design methodology \citep{Suresh2021-qq, Dhanorkar2021-ah}. 
During this phase, we also conducted a supplementary search to identify additional relevant articles using both forward and backward citation tracking.
}

\added{
The technical surveys include reviews of interpretable algorithms in general \citep{Linardatos2020-ay}, from a neural-symbolic standpoint \citep{Townsend2020-vv}, through knowledge graphs \citep{Tiddi2022-dc}, or within specific domains, such as CyberSecurity \citep{Capuano2022-gy} and Medicine \citep{Sheu2022-al}. 
While we do not examine these technical papers in detail, they illustrate how AI and HCI communities employ different taxonomies and categorization schemes for XAI.
AI researchers typically categorize XAI by explanation scope or algorithmic characteristics, whereas HCI researchers emphasize user outcomes and interaction modes.
We provide a more comprehensive overview of technical XAI concepts and taxonomies in Appendix \ref{appendix:XAI terminologies} and outline the different types of explanations in Appendix \ref{appendix:types of AI explanations}.
}

\subsubsection{Phase 2: Screening Empirical Literature and Mapping Rhetorical Effects}
\added{
Group II comprises empirical studies that document XAI design processes and report the effects of AI explanations on users. 
This focus allows us to link specific XAI design features to quantitative measures and qualitative feedback, enabling critical reflection of how design choices manifest rhetorical effects in concrete user tasks.
} 

\added{
Because rhetoric remains under-examined in XAI research, there is no established procedure for determining whether reported outcomes robustly demonstrate rhetorical effects. 
As a result, mapping XAI design strategies to rhetorical appeals necessarily involved interpretive judgment based on the authors' expertise in design, HCI, and AI. During this process, we excluded papers that lacked sufficient detail (to reliably identify rhetorical effects), diverged from rhetorical assumptions, or fell outside the scope of our analysis. Additional rationales and examples of excluded papers are described in Appendix~\ref{appendix:phase 2}. We provide more details on this mapping process in Section \ref{sec:Framework}, where we introduce the rhetorical framework.   
}




\subsection{Reflexivity Statement} \label{subsec:reflexivity}
\added{
Our research team includes experts in design, HCI, and AI. 
The initial framing of this work drew heavily on the primary author's academic background in communication and design \citep{Buchanan1985-qy}, with a particular focus on visual rhetoric \citep{Foss2004-hu}. 
This naturally led us to conceptualize XAI through a communicative perspective (i.e., perceiving explanations as designed artifacts rather than solely technical products) and to underscore the limited integration of rhetorical theory in prior XAI research.
Team members with AI expertise contributed technical perspectives on how current-state-of-the-art interpretable models provide faithful rationales for algorithmic behavior. 
Through this interdisciplinary collaboration, design and AI researchers jointly examined how machine logic is translated (or not) into design representations.
}

\added{
All authors contributed to the development of the proposed rhetorical framework (Section~\ref{sec:Framework}).
Two authors, a design specialist and an AI expert, conducted the two-phase literature screening. 
We acknowledge that our disciplinary backgrounds informed all stages of the narrative review, including keyword selection, screening decisions, framework construction, and interpretation of design strategies.
To mitigate interpretive bias, coding decisions were iteratively discussed among the authors and grounded in established theories from rhetorical HCI (Section \ref{subsubsec:rhetorical theories}). 
Accordingly, the rhetorical strategies and user effects presented in Section~\ref{sec:Results} are positioned as a narrative summary, rather than as an exhaustive or prescriptive account.
}

\section{Existing XAI Design Paradigms and The Rhetorical Complementarity} \label{sec:XAI design}
Our narrative review yields three XAI design paradigms, namely user-centered (Section \ref{subsec:user-centered XAI}), stakeholder-centered (Section \ref{subsec:stakeholder-centered XAI}), and socio-technical (Section \ref{subsec:sociotechnical XAI}). \added{Based on foundationals in rhetorical HCI  (Section \ref{subsec:rhetorical HCI}), we also describe how rhetoric complements and extends these design paradigms (Section \ref{subsec:complementory}).}

\subsection{User-centered Explanations} \label{subsec:user-centered XAI}

HCI scholars studying XAI have focused on how different types of explanations can support users in understanding AI systems more effectively \citep{Hagras2018-hx, Liao2020-da, Vera_Liao2021-zi}. \citet{Hagras2018-hx} first argued that XAI aimed to \textit{make AI understandable to humans}. One strategy toward this goal is to reframe technical descriptions of AI decision-making into user-friendly language. For example, consider explaining  how an autonomous car changes speeds. Instead of a technical explanation such as ``if the distance to the car ahead is less than 2.5m and the road is 10\% slippery then reduce car speed by 25\%,'' a more human-like natural language rephrase might be ``if the distance to the car ahead is low and the road is slightly slippery, then slow down.'' In general, researchers have developed various types of explanations (see Appendix \ref{appendix:types of AI explanations}) that employ visualizations and natural language to make AI more accessible and understandable to people \citep{Lai2023-bu, Robbemond2022-px}.

Given the diversity of explanations, researchers have proposed different frameworks to help designers select the most appropriate explanation type for a given context. For example, \citet{Liao2020-da} proposed an XAI ``question bank'' comprised of common user inquiries such as \textit{How} AI works, \textit{Why} AI made a specific prediction, \textit{Why not} an alternative prediction, \textit{What} AI would predict if the input changes, and \textit{How} to change the input to alter the AI prediction. Other researchers have adopted this question-informed framework to select AI explanations when designing XAI systems \citep{Xin_He2023-dj, Ehsan2022-fz, Wang2023-pk}. Other frameworks aim to bridge human understanding and machine explanations from a cognitive perspective. Examples include \citet{Wang2019-we} proposing to design AI explanations align with human cognitive processes, \citet{Chen2022-cq} incorporating task-specific human intuitions when assessing user understanding of explanations, and more \citep{Zhang2019-wj, subhash2022makes, Chen2023-me}.

Since AI systems serve diverse end-user groups, explanations can be tailored to their specific needs and contexts \citep{Ehsan2024-lu, Kim2024-hp}. For example, loan seekers and lenders have different explanatory needs for understanding how a credit scoring AI system functions \citep{Bove2022-pd}. Loan seekers might be particularly interested in why their financial record led to a rejection and how to alter the decision. Lenders might seek a global understanding of AI behavior across historical loan records to support financial analysis, enabling them to make more informed decisions about issuing loans to future applicants \citep{Dikmen2022-hd}. Similar distinctions are also prevalent in the medical domain between patients and healthcare providers \citep{Burgess2023-dm, Kim2024-hp}. Collectively, these cases highlight how domain expertise shapes explanatory needs \citep{Mohseni2021-bw, Kim2024-hp}. Lay users often prefer personalized and actionable guidance, while experts commonly seek system-level understanding.

\subsection{Stakeholder-centered Explanations} \label{subsec:stakeholder-centered XAI}

While end-users are a central focus in XAI, they represent just one of the many stakeholders involved in developing, deploying, and adopting AI systems. To situate AI within this broader socio-technical ecosystem, explanations should also cater to stakeholders' distinct goals, responsibilities, and information needs. Two primary approaches for categorizing stakeholders are based on their \textit{roles} within the AI lifecycle \citep{Tomsett2018-ij, Preece2018-ql, Hoffman2021-ek, Dhanorkar2021-ah} and their level of \textit{AI knowledge} \citep{Mohseni2021-bw}.

In role-based stakeholder categorization, \citet{Hoffman2021-ek} demonstrated distinct explanatory needs between developers, policymakers, and domain users. They reported that developers mainly employed explanations to debug AI errors and identify potential algorithmic biases to improve overall predictive performance. Policymakers leveraged explanations to audit AI behavior in expected deployment settings and unexpected edge cases. Domain users used explanations to understand or counterfactually remediate unfavorable AI decisions. Other researchers also used role-based frameworks but with alternative groupings of stakeholders \citep{Tomsett2018-ij, Preece2018-ql, Dhanorkar2021-ah}. 

Complementary to stakeholder roles, \citet{Mohseni2021-bw} proposed a taxonomy of explanation needs based on varying levels of AI expertise. They distinguished between \textit{AI novices} (layman users) from \textit{AI experts} (engineers and developers), each leveraging explanations to suit different purposes. 
However, \citet{Suresh2021-qq} contended that both role-based and AI knowledge-based frameworks could be problematic, as overlapping explanation needs could challenge consistency between different stakeholder groups.
Thus, they proposed to organize stakeholder needs around \textit{Goals}, \textit{Objectives}, and \textit{Tasks} along the AI development, aiming to offer a more cohesive and ontologically grounded vocabulary. 

While stakeholder frameworks might differ in structure and granularity, they all highlighted the need to contextualize explanations to specific goals and responsibilities of different stakeholders \citep{Wolf2019-sv, Carroll2003-uw}. Identified stakeholder needs then served as specifications to guide the technical development \citep{Kaplan2024-td} and interface design of XAI systems \citep{Dhanorkar2021-ah, Ehsan2023-pu}.

 
\subsection{Socio-technical Explanations} \label{subsec:sociotechnical XAI}

As researchers expand their focus from end-users to a broader set of stakeholders, 
\added{
attention shifts toward organizational and infrastructural challenges beyond what user-tailored explanations alone can address, reflecting a socio-technical turn in XAI \citep{Preece2018-ql, Ehsan2020-gz, Dhanorkar2021-ah}.
}
\deleted{this shift reflects a socio-technical perspective for studying XAI}. 
\citet{Brandao2019-qi} argued that AI experts often lack a natural understanding to connect the potential social meanings with models they create. For example, \citet{Ehsan2023-pu} highlighted a sales prediction tool with 90\% accuracy that achieved only 10\% adoption. While inexperienced sellers over-trusted it, experienced sellers disregarded it entirely because it didn't align with their professional intuition. Similar cases have emerged in different areas, including customer-service relationship \citep{Northey2022-it}, journalist norms and values \citep{Komatsu2020-xz}, and frontline workers \citep{Fox2023-gy}. The core issue \deleted{described in the above use cases} is that current AI deployment did not account for organizational factors \added{such as power dynamics, workflows, and culture}. 
\added{
In particular, users reported that working with AI often added invisible organizational work rather than facilitating their tasks, including learning new terminology, consulting technical experts, and managing transparency concerns \citep{Ehsan2021-id, Ehsan2023-pu}.
}
\deleted{To address this, some XAI researchers argued that explanations should also account for user roles, expertise, and collaborative practices within the organization}.

\subsection{The Complementary Role of Rhetorical XAI} \label{subsec:complementory}

Thus far, previous sections have outlined key design paradigms of human-centered XAI, including user-centered (\ref{subsec:user-centered XAI}), stakeholder-centered (\ref{subsec:stakeholder-centered XAI}), and sociotechnical perspectives (\ref{subsec:sociotechnical XAI}).
\added{
Across these paradigms, explanation design is consistently grounded in (1) \textbf{a need-driven rationale} that derives explanatory goals from users or stakeholders, and (2) \textbf{a cognitive alignment} that seeks to mirror human reasoning in AI explanations.
These rationales also shape how trust and reliance are conceptualized and evaluated in XAI, positioning them as explanatory goals achieved by satisfying identified user needs or by aligning human and AI mental models in decision-making.
For example, the notion of \textit{calibrated trust} is operationalized through prescriptive measures that assess whether users appropriately accept correct AI decisions and reject incorrect ones, reflecting an implicit assumption that trust requires an accurate understanding of model accuracy \citep{Lee2004-kt, Zhang2020-pf, Wang2021-hv}. 
Similarly, \textit{reliance patterns} such as automation bias and algorithmic appreciation are defined and measured using comparable criteria, treating appropriate reliance as evidence that users’ decision strategies are aligned with the AI system’s capabilities and failure modes \citep{Cabitza2023-ts, Cao2024-ed}.
}

\added{Inspired by rhetorical HCI (Section~\ref{subsec:rhetorical HCI}),} we ask, \textit{what if designers instead see our roles as rhetoricians?} This perspective extends beyond an emphasis on logical rationales alone and highlights the interplay among \textit{logos}, \textit{pathos}, and \textit{ethos} in shaping how AI explanations are constructed and interpreted. We identified two complementary insights from the existing logical XAI foundations. 

\added{
\textbf{Insight I: Rhetoric motivates us to explore what makes AI systems persuasive and compelling for use.}
Scholars in HCI \citep{Churchill2020-mg} and Design \citep{Friess2010-tf} have long noted a translational gap from research to practice. 
In real-world contexts, users may not strictly adhere to needs identified in research \citep{Norman2005-pz, Buchanan1992-bo} or engage in fully rational decision-making assumed in pre-defined scenarios \citep{Kahneman2012-cs, Suchman1987-oq, Norman2006-sk}.
} 
In XAI, understanding how AI works (i.e., pure interpretability) is the \textit{means} rather than the \textit{end} goal for why people choose to use AI in the first place. \deleted{This echoes related critiques in the XAI literature.} For example, \citet{de-Bruijn2022-wh} observed that ``Explanations are situationally dependent and can change over time.'' Likewise, \citet{Miller2021-nt} claimed that ``Explanations of how models work might just distract people from figuring out what they really need or want to know.'' \deleted{Our call to broaden XAI resonates with prior work, which argued that effective use of AI may not require a deep understanding of model behaviors. Therefore, explanations should go beyond the current focus on functional transparency, addressing more rhetorical questions such as: \textit{What is the compelling value proposition for a user to engage with AI?}}

\added{
People use AI for various reasons. 
Some engage with AI to meet organizational or legal obligations, others to accelerate their work, others to support exploration and creativity, and others for personal enjoyment.
These diverse motivations can be analyzed through the rhetorical appeals of \textit{ethos}, \textit{logos}, and \textit{pathos}, extending beyond simply understanding how AI works to also justifying why AI merits use.
By foregrounding different rhetorical dimensions, rhetoric supplements need-based rationales that design explanations primarily around articulated user needs \citep{Friess2010-tf, Norman2006-sk}, as well as cognitive formulations of trust that treat appropriate AI use as a function of understanding model behavior.
This expanded perspective enables us to account for why certain designs thrive in contexts that contradict rational assumptions and, conversely, why certain designs fail even when rational needs appear to be satisfied.
}

\added{\textbf{Insight II: The effectiveness of an explanation is not an intrinsic property, but a situated outcome of the rhetorical triangle: an interplay between the explanation’s source, its message, and the recipient’s role under a specific task context.} 
This perspective helps remediate seemingly ``contradictory" results in empirical studies.
For example, \citet{Dodge2019-ss} examined how different types of explanations influenced human fairness evaluations of an ML system. 
They found that case-based explanations (showing how many individuals in the training set were identical to the test case) were perceived as less fair than other explanation types.
In contrast, \citet{Chen2023-vc} found that case-based explanations better supported appropriate reliance by aligning with user intuitive reasoning about outcomes, features, and model limitations.
These differing effects of case-based explanations reflect how explanation effectiveness emerges from distinct rhetorical configurations of task, audience, and evaluative focus.
\citet{Chen2023-vc} primarily evaluated participant perceived logical validity of AI explanations in tasks such as income prediction and biography classification.
On the other hand, \citet{Dodge2019-ss} situated explanations within a high-stakes criminal justice context, where fairness judgments are embedded in moral accountability and institutional credibility, rendering case-based explanations less persuasive.}

\vspace{0.2cm}

\noindent
\added{In summary, we argue that designing appropriate AI explanations for a given context requires careful consideration of the explanation source, its intended message, and its target recipients.
Accordingly, designers should avoid evaluating explanation effectiveness solely in terms of perceived logical clarity or need-based rationales.
Instead, explanation design should account for the situated rhetorical effects that emerge from the interplay of: \textit{logos}, the alignment of technical logic with human reasoning based on the visual and textual abstract representation; \textit{ethos}, the contextual credibility established by the explanation source and its appropriateness for the decision task; and \textit{pathos}, the emotional resonance achieved by addressing the user motivations, expectations, or situated needs during interaction.} 

\section{A Rhetorical Framework and Coding Procedure for Identifying Rhetorical Strategies} \label{sec:Framework}

\begin{table*}[ht]
\resizebox{\textwidth}{!}{%
\begin{tabular}{@{}p{0.12\textwidth}>{\raggedright}p{0.4\textwidth}>{\raggedright\arraybackslash}p{0.4\textwidth}@{}} \toprule
\multirow{2}{*}{Dimensions} & \multicolumn{2}{c}{Rhetorical Inquiries for AI Explanations}  \\ \cmidrule(r){2-3}
& \multicolumn{1}{c}{\textbf{Explain How AI Works}} & \multicolumn{1}{c}{\textbf{Explain Why AI Merits Use}} \\ \midrule
\textbf{Logical} \newline \textbf{Reasoning}  &  What forms of explanations are used, and how are they presented to convey the underlying logic of AI local prediction or overall behavior? &  How do these explanations help end-users apply AI effectively to address their real-world needs and tasks? \vspace{0.2cm} \\  
\textbf{Credibility}   &  How does the framing and presentation of explanations signal their credibility to users? &  How do the framing and presentation of explanations signal the credibility of AI systems? \vspace{0.2cm} \\        
\textbf{Emotional Resonance}  &  How do the framing and presentation of explanations engage user emotions to facilitate their understanding of AI? & How do the framing and presentation of explanations engage user emotions to better utilize the AI system? \\  \bottomrule
\end{tabular}
}
\vspace{0.2cm}
\caption{A Rhetorical Inquiry Framework to Characterize AI Explanations.}
\label{tab:rhetorical AI explanations}
\end{table*}

Our framework builds on the two insights detailed in Section~\ref{subsec:complementory}. \deleted{as outlined in the previous section.} \deleted{Drawing on the first idea --- AI explanations function as rhetorical arguments, motivating us to explore what makes AI explanations persuasive and compelling for use ---}First, we propose extending XAI beyond merely (1) explaining \textit{How an AI system works} \deleted{(e.g., clarifying local predictions and global behaviors)} to also (2) explaining \textit{Why an AI system merits use}. Next, \deleted{we apply the second idea --- strategies can be identified from the interplay around three unique rhetorical triangle --- to guides} \added{we apply the rhetorical triangle to guide} our synthesis of rhetorical strategies in prior XAI literature. By aligning these three dimensions with key actors emphasized in different human-centered XAI paradigms, we can uncover a range of rhetorical strategies that operate across multiple levels.


Building on this conceptual adaptation, rhetorical design serves two complementary objectives in XAI: (1) \textbf{supporting traditional AI explanations that clarify \textit{How AI Works}}, and (2) \textbf{designing explanations that communicate \textit{Why AI Merits Use} to guide or inform adoption}. Our goal in foregrounding this distinction is not to enforce a rigid dichotomy, but to highlight the nuanced differences between these design objectives and to propose corresponding strategies. More importantly, we hope that clarifying both objectives and keeping them in mind during the design process can enhance the value of explanations, suggesting which types of explanations to use and the contexts where they are appropriate or inappropriate regarding the technology affordance, task requirements, and the stakes involved (we discuss this further in Section \ref{subsec:task relevance}). 



We examine each objective through the three rhetorical dimensions: logical reasoning, credibility, and emotional resonance.
In Table \ref{tab:rhetorical AI explanations}, we outline each rhetorical appeal as a coding inquiry across the two explanatory goals. While Table \ref{tab:rhetorical AI explanations} contrasts the two goals to clarify our coding process, this framing is intended to support, not discourage, explanation strategies that span both objectives. \added{In the following subsections, we adapt the open and axial coding from \citet{Corbin2008-rl} into a structured three-step procedure guided by our rhetorical framework.} 

\subsection{Step 1: Classifying Design Examples based on XAI Objectives} Before applying open coding to identify specific design strategies, we first conducted a top-down process to classify prior work based on whether AI explanations were primarily used to (1) convey how the AI system works or (2) 
enhance the usability and adoption of the AI. 
Specifically, we conducted a need-based meta-analysis grounded in the principles of user-centered XAI (Section \ref{sec:Background}). 

For example, \citet{Cai2019-lb} presented example-based explanations to help users understand how an AI model recognizes a drawing of an avocado. 
Their primary motivation was to ``help users develop intuition for the reasons underlying machine predictions, without directly exposing the internal logic of the algorithm.''
Therefore, we classified \citet{Cai2019-lb} under the first XAI objective: explaining how AI works. 
In contrast, in another work by \citet{Cai2019-zx}, the authors integrated example-based explanations and other interactive features into a case-based retrieval system to inform clinical diagnosis. We classified \citet{Cai2019-zx} under the second XAI objective since the presented explanations primarily demonstrate the practical benefits of the AI system. Additionally, we classified \citet{Wang2023-ar}'s interactive recourse tool for loan approval under both XAI goals, as explanations provided not only support user understanding of model behavior but also offer actionable insights to help users navigate the application and address real-world challenges.

We further validated our classification by examining the AI expertise of recruited participants and the relevance of the AI task to their professional work. As a rule of thumb, studies involving domain experts were more likely to pursue the second objective, since specialists could more easily evaluate the utility of XAI systems in real-world contexts.
Conversely, crowdsourced studies were more likely to focus on the first objective, as the tasks were typically less grounded in the specific needs and contexts of real-world users. Of these, we flagged studies that employed attention checks or screening criteria to recruit participants with relevant domain knowledge.

\subsection{Step 2: Identifying Design Strategies} During open coding, we first collected and organized prior XAI systems based on different types of explanations (Appendix \ref{appendix:types of AI explanations}). 
Next, we scrutinized the presentation of explanations, focusing on design components such as visual elements (e.g., lines, shapes, and graphics) and linguistic features (e.g., languages and tones). 
We inductively derived an initial set of codes characterizing recurring design strategies from this process. 

\subsection{Step 3: Associating Design Strategies with Rhetorical Appeals} 
Finally, we conducted axial coding to organize recurring design patterns by aligning them with the three rhetorical appeals: logical reasoning, credibility, and emotional resonance. We analyzed reported quantitative and qualitative user feedback to assess how design strategies influenced user perceptions. This step connected design intent to user impact by examining how specific strategies contributed to rhetorical persuasiveness.

For example, for quantitative measurements, ``simulatability'' (i.e., participants predicting AI behavior) and ``perceived usefulness'' assess the effects of logical appeals, as they reflect improvements in user understanding of AI reasoning and AI utility. For credibility appeals, ``self-confidence'' and ``perceived trust'' indicate how trustworthy the AI system appears. Similarly, for emotional appeals, ``affective reaction scores'' and ``user engagement metrics'' quantify the extent to which explanations resonate affectively with users. 

Additionally, we contextualized quantitative measures with reported qualitative feedback. 
We present the reviewed articles and their corresponding coding schemes in \textbf{Table~\ref{tab:coding scheme}}. For each paper, we specified the primary XAI goal, identified strategies, and mapped both quantitative and qualitative feedback to the appropriate rhetorical appeals. Each code assigned to an excerpt from the articles indicates the design intent it represents and the rhetorical appeal it conveys.


However, we acknowledge that our meta-analysis is selective rather than comprehensive. For example, in some cases, design strategies identified from papers primarily targeting the first XAI goal could also be applicable to the second goal. Moreover, not all user studies in prior work captured responses across all three rhetorical dimensions. 

\section{Rhetorical Strategies} \label{sec:Results}

\added{As shown in Table \ref{tab:rhetorical_strategies},} we categorize identified strategies based on three rhetorical appeals: \textit{Logical Reasoning} (Section \ref{subsec:Logos}), \textit{Credibility} (Section \ref{subsec:Ethos}), and \textit{Emotional Resonance} (Section \ref{subsec:Pathos}). 
Within each appeal, we describe how different design strategies support the two XAI objectives: explaining how AI works or why AI merits use (see Table~\ref{tab:rhetorical AI explanations}). 
We assess each strategy's strengths and limitations based on axial coding process, linking reported user experiences to their intended rhetorical appeals. 

\begin{table*}[ht]
\centering
\begin{tabular}{@{}lll@{}} 
\toprule
\multirow{2}{*}{Dimensions} 
& \multicolumn{2}{c}{Rhetorical Inquiries for AI Explanations} \\ 
\cmidrule(lr){2-3}
& \textbf{Explaining How AI Works} 
& \textbf{Explaining Why AI Merits Use} \\ 
\midrule
\makecell[l]{\textbf{Logical} \\ \textbf{Reasoning}} 
& \makecell[l]{Isolated Presentation (\ref{sec:isoPre}) \\ Adaptive \& Mechanistic Presentation (\ref{sec:adapPre}) \\ Graphical Synergy (\ref{sec:synergy})} 
& \makecell[l]{Explanations as Recommenders (\ref{sec:expRec}) \\ Explanations as Retrievers (\ref{sec:expRet})} \\  
\textbf{Credibility} 
& \makecell[l]{Authority Framing (\ref{sec:authFrame}) \\ Cultural Relatedness (\ref{sec:culturalRel}) \\ Numerical Association (\ref{sec:NumerAsso})} 
& \makecell[l]{Technical Assurance (\ref{sec:TechAssur}) \\ Social Endorsement (\ref{sec:socEnd})} \\        

\makecell[l]{\textbf{Emotional} \\ \textbf{Resonance}} 
& \makecell[l]{Explanation Affect (\ref{sec:exaAffect}) \\ Emotional Cues (\ref{sec:emoCue}) \\ Protective Framing (\ref{sec:protFram})} 
& \makecell[l]{Expectation Management (\ref{sec:expMan}) \\ Expressive Interaction (\ref{sec:exprInte})} \\  
\bottomrule
\end{tabular}
\vspace{0.2cm}
\caption{A taxonomy of rhetorical strategies organized by two explanatory goals: How AI works and Why AI merits use.}
\label{tab:rhetorical_strategies}
\end{table*}

\subsection{Logical Reasoning} \label{subsec:Logos}

A key focus of XAI is explaining the logical reasoning behind automated decisions to foster appropriate trust in AI predictions or align user mental models with overall AI behaviors \citep{Wang2019-we, Miller2019-zh, El-Assady2022-xa}.
We identify three design strategies from prior XAI research: \textbf{Isolated Presentation (\ref{sec:isoPre})}, \textbf{Adaptive and Mechanistic Presentation (\ref{sec:adapPre})}, and \textbf{Graphical Synergy (\ref{sec:synergy})}. 
Explanations can also communicate why AI merits use by positioning the AI as an intelligent, efficient, and rational collaborator. Toward this end, we identify two design strategies that directly enhance the usability and adoption of an AI system: \textbf{Explanations as Recommenders (\ref{sec:expRec})} and \textbf{Explanations as Retrievers (\ref{sec:expRet})}.


\subsubsection{Isolated Presentation} 
\label{sec:isoPre}
Different types of AI explanations (see Appendix \ref{appendix:types of AI explanations}) inherently embody distinct forms of human logical reasoning. \citet{Wang2019-we} conceptually mapped how various machine learning models reason and generate explanations to corresponding forms of human decision-making. For example, they argued that Bayesian probability represents inductive reasoning, a bottom-up approach that uses prior intuitions to draw broader conclusions. 
Case-based models encapsulate analogical reasoning, which solves new problems by synthesizing solutions from previously solved problems having similar issues. 
Rule-based models illustrate deductive reasoning, a top-down approach that begins with general principles and operationalizes them to specific cases.

Explanations can be compelling logical arguments when their embodied reasoning aligns with user mental models. When explaining how AI works, the presented explanation reinforces trust and interpretability if users perceive the mapping from inputs to outputs as logically coherent or aligned with their expectations. 
In summary, different types of AI explanations operationalize different modes of human reasoning, and their persuasiveness depends on how well embodied logic aligns with user expectations.

A dominant design strategy employed in prior XAI studies is \textbf{isolated presentation}, where different types of explanations are presented individually for A/B testing. 
Specifically, these studies examined the extent to which distinct logical structures embodied in different explanations (e.g., features, examples, rules, or concepts) helped to improve human-AI complementary performance \citep{Bansal2020-zw, Gajos2024-la, Nguyen2018-az} or promote appropriate reliance on AI predictions \citep{Schemmer2023-zy, Yang2020-wm, Zhang2020-pf, Liu2021-ds, Vasconcelos2023-cm}.
For example, in comparing feature-based and example-based explanations, \citet{Chen2023-vc} observed that the latter elicited more critical reflections on potential misalignments between human intuition and machine reasoning. Moreover, \citet{Van_der_Waa2021-he} demonstrated that rule-based explanation showed a small positive effect on (global) system understanding compared to example-based explanation. Nevertheless, both types of explanations could lead to confirmation bias \citep{Bertrand2022-xg}, where users tended to over-rely on AI predictions erroneously. 


\added{
A prominent issue in XAI centers around faithfulness, the challenge of ensuring that an AI's explanation accurately reflects its true internal reasoning \citep{jacovi-goldberg-2020-towards}. While faithfulness is often framed as a technical property, design choices also shape how faithfully model behavior is communicated to users through different explanatory strategies.
}
Regardless of format or style, all AI explanations are paraphrased distillations of the underlying computational algorithm. Isolated evaluations of explanations may simplify analysis, but can also mislead users by reinforcing prior (erroneous) assumptions or painting an incomplete picture of system behavior \citep{El-Assady2022-xa}. These risks arise in part because AI systems and the explainable techniques used to interpret them (described in Appendix \ref{appendix:XAI terminologies}) are not inherently designed to align with human reasoning \citep{Chen2023-me, Wang2019-we}. Although neural networks could be loosely modeled on the structure of the    human brain \citep{Schaeffer2022-jo}, AI systems are typically trained to rely on complex, uninterpretable features that differ fundamentally from the semantic reasoning humans use to understand the world  \citep{Fei-Fei2007-lm, Zhang2019-wj}. As a result, explanations may appear plausibly persuasive to users, but do not necessarily reflect the actual decision-making process of the system or reliably align with human reasoning.

Furthermore, explanations may be misinterpreted when subtle variations in visual elements convey unintended or oversimplified meanings \citep{Lupton2015-lg, Gibson1899-hv}. 
For example, \citet{Ye2023-zy} observed that subtle differences in saliency maps design could significantly influence annotator perceptions in uncertainty-aware medical semantic segmentation. They compared three formats (singular highlights, continuous gradient maps, and confidence contour maps) and found that the first two were less interpretable. In particular, continuous gradients were prone to confusion, as annotators struggled to infer clear thresholds for distinguishing uncertainty levels. Other studies reported similar qualitative feedback, where participants noted that continuous gradient maps of local feature importance only helped recognize relevant features but difficult to quantify their relative importance \citep{Alqaraawi2020-ns}. These failures are also observed in textual contexts \citep{Schuff2022-ly}.

Based on extensive prior research examining the user effects of various explanations, isolated explanations \added{are employed} \deleted{may be most suitable} in early-stage design and evaluation settings, where the goal was to investigate how different explanation formats align with the needs \citep{Liao2020-da} and preference of a target stakeholder group \citep{Dhanorkar2021-ah, Panigutti2023-fm, Tonekaboni2019-jo}. This approach is also \added{used to understand} \deleted{effective for understanding} user mental models in controlled experiments, such as identifying the error boundaries of human-AI decision-making \citep{Bansal2019-ig, Bansal2019-lw}. Notably, isolated explanations provide a clear basis for comparison before progressing to more advanced designs, such as adaptive and mechanistic presentation, or graphical synergy, which reveal AI behaviors in greater complexity (as described in the following sections).

\subsubsection{Adaptive and Mechanistic Presentation} \label{sec:adapPre}
Since AI systems do not inherently follow human reasoning, presenting isolated explanations risks reinforcing misconceptions by oversimplifying AI behavior.
To mitigate this, researchers have principally adopted two design strategies. \textbf{Adaptive presentation} dynamically adjusts explanation format based on evolving user objectives or deployment contexts. \textbf{Mechanistic} presentation combines multiple explanation types into an integrated \textit{UI dashboard}.
%
Both extend isolated presentations through adaptive adjustments or mechanical aggregations. While the mode of delivery differs, the underlying logical appeals of these explanations remain unchanged. This contrasts with synergistic design strategies introduced later (Section \ref{sec:synergy}), which fundamentally reconfigure explanatory logic. 

Adaptive explanations are primarily delivered in two ways: (1) through user requests, such as in a conversational setting, or (2) automatically triggered through internal system indicators. 
For user-initiated requests,  \citet{Miller2019-zh} showcased an idealized explanation dialogue in which an XAI agent provides different types of explanations (e.g., features, rules, examples) in response to diverse user inquiries on how the AI classifies an input image as a spider (versus a beetle or octopus). 
\citet{Shen2023-kk}  operationalized this adaptive dialogue in human-AI scientific writing contexts. They constructed a chatbot to provide heterogeneous AI explanations across different stages of human-AI communication.  

For system-triggered deliveries, explanations are automatically adjusted in response to internal (invisible to end-users) indicators. For example, some XAI systems present explanations only when AI confidence is low \citep{Bansal2020-zw} or dynamically adjust explanations based on real-time accuracy updates or error boundaries \citep{Bansal2019-lw, Bansal2019-ig, Wang2023-qb}. Others adapt explanations based on human learning progress \citep{Wambsganss2020-fq} or show explanations only when the AI correctness is estimated to exceed that of the user \citep{Ma2023-rj}.
Compared to adaptive presentations, mechanistic presentations of explanations are generally more straightforward to implement since no real-time adaptations are required. 
For example, \citet{Wang2019-we} integrated local feature importance, global feature contributions, and counterfactual rules into an AI-assisted medical diagnosis tool, enabling clinicians to consult multiple explanation types simultaneously when evaluating patient cases. 
Similarly, \citet{Bove2022-pd} developed an explainable auto-insurance interface that helped vehicle owners understand how their driving behaviors influenced insurance premiums. Likewise, \citet{Cau2025-rj} created an explainable hiring dashboard that enabled recruiters to accept or reject applications based on applicant attributes through multiple types of explanations.

Overall, both adaptive and mechanistic presentations offer unique benefits to improve user understanding and support human-AI teaming. Adaptive presentations enhance relevance by tailoring explanations to user goals, expertise, or decision context, reducing cognitive overload and aligning responses with moment-to-moment needs. \deleted{Thus, adaptive explanations are particularly valuable in dynamic or iterative workflows, where user needs may evolve \citep{Suresh2021-qq}.}
Mechanistic presentations offer breadth and transparency by juxtaposing multiple explanation types in a unified interface, allowing users to cross-reference local and global reasoning or compare factual and counterfactual views.
\deleted{As a result, they are more effective in diagnostic or investigative tasks, where users benefit from cross-checking different explanatory lenses to reach more robust conclusions \citep{Miller2023-nc}.}
Compared to isolated presentations, \textit{interactivity} is a crucial differentiating factor for adaptive and mechanistic presentations, allowing logical reasoning processes to unfold dynamically in response to user actions or system feedback. 

Despite their benefits, \citet{Gaole2025-wo} found that both adaptive and mechanistic presentation formats could lead to a clear over-reliance on the AI system. The authors attributed this over-reliance to \textit{an illusion of explanatory depth} where the presence of multiple explanations, regardless of actual fidelity, reduced critical evaluation of system behavior. This echoes a classic rhetorical strategy in communication, where shifting the burden of proof makes arguments appear more persuasive \citep{Hahn2007-pv}.

\subsubsection{Graphical Synergy} \label{sec:synergy}
Regardless of whether explanations are presented in isolated, adaptive, or mechanistic formats, different explanation types often coexist unilaterally without being cohesively integrated. 
As \citet{Wang2019-we} noted, this fragmentation arises because the underlying explanatory algorithms rely on fundamentally different logical bases.
Nevertheless, emerging work has shown promise in organically merging diverse explanation types into a visually coherent representation, an approach we term \textbf{graphical synergy}. Due to its visual nature, graphical synergy can be effective for pattern recognition tasks, where synthesizing complementary explanation types helps users grasp subtle relationships that no single explanation could convey.

For example,  \citet{Evans2022-zg}  synergistically combined normative example-based explanations with counterfactual explanations, forming a cohesively contrastive causal chain.
Specifically, they presented prototypical examples alongside counterfactual instances of cell nuclei to help clinicians differentiate between positive and negative Ki67 scores used to estimate tumor proliferation. 
Compared with traditional saliency maps, prototypical examples, and concept attributions, clinicians reported that the synergistic presentation ``help[ed] them better understand what the algorithm is looking for... it is self-explanatory that staining was the most important factor distinguishing positive nuclei from those marked negative, with some identifying other important factors such as shape [and] size of nuclei."
Clinician subjective ratings reinforced this feedback, where the synergistic explanation format received the highest median scores for clarity, relevance, trustworthiness, and practical utility.

While graphical synergy can improve explanation coherence, integrating multiple explanation types can raise concerns about fidelity to the underlying AI model. 
To synthesize transitional examples between prototypical positive and negative Ki67 nuclei instances, \citet{Evans2022-zg} combined data attribution methods with image morphing techniques \citep{Gulshad2021-hv}.
Because these synthetic interpolations are not grounded in training data, they raise faithfulness concerns that highlight the need for new XAI techniques that seamlessly integrate explanation types while preserving fidelity.
Developing such techniques would support richer graphical synergies, such as combining example-based explanations with saliency highlights \citep{Kenny2021-ts} or integrating salience maps with human-interpretable concepts \citep{Yang2024-DC}. 


\subsubsection{Explanations as Recommenders} 
\label{sec:expRec}
Traditionally, XAI work focused on explaining how systems work by helping users to assess, replicate, or predict AI decisions. 
Emerging work now shifts toward explaining why AI systems merit use, particularly by supporting users in solving real-world problems beyond simple classification. 
In these contexts, explanations act as \textbf{recommenders}, helping users understand what actions to take and why those actions are beneficial.

For example, \citet{Wambsganss2020-fq} developed an adaptive AI-assisted learning system to help students  write persuasive arguments. 
To facilitate revisions, the system visualized argumentative structure, evaluated writing quality, and provided natural language feedback alongside highlighted areas for improvement.
Students expressed a significantly greater preference for using the new system than traditional reference-based writing tools.
\citet{Nilforoshan2018-cc} implemented a similar application using feature-based explanations to help reviewer draft social media responses. 

Besides writing, algorithmic recourse represents another practical use case of explanations. 
For instance, \citet{Kim2024-hp} developed an XAI interface that combined counterfactual examples with recourse plans to help dysarthria patients correct mispronunciations.
Similarly, \citet{Wang2023-ar} introduced \textit{GamCoach}, which provided loan applicants with counterfactual explanations recommending financial changes to improve loan approval odds.

Findings from \citet{Wang2023-ar} highlight a key point that aligns with our framework (outlined in Section \ref{sec:Framework}): \textbf{effective use of an AI system does not necessarily require a deep understanding of system behaviors}. 
For example, in their user study, participants rated ``ease of understanding'' (4.90) and ``help with understanding the model'' (5.17) lower than their ``intention to reuse the system for future loan applications'' (5.34) and ``the system’s usefulness in identifying ways to improve their loan applications'' (5.61), based on a 7-point Likert scale.
Although observed differences were not statistically validated, this finding suggests that a deep understanding of model behaviors is not a prerequisite for effectively using the AI for algorithmic recourse. 

\deleted{To summarize, explanations as recommenders are particularly beneficial in scenarios where users seek actionable guidance. By positioning AI as a collaborative advisor, this strategy reinforces our position on XAI's rhetorical role: explanations should reveal how AI works and why an AI merits use for practical utility.}


\subsubsection{Explanations as Retrievers} \label{sec:expRet}
Explanations can also function as \textbf{retrievers} to support dataset exploration and analogical reasoning. 
Different types of explanations highlight distinct model signals (see Appendix~\ref{appendix:types of AI explanations}), including local feature attributions, global patterns, representative prototypes, and human-interpret concepts. 
Users can leverage these characteristics to retrieve analogous cases, facilitate structured comparisons along shared attributes, and derive actionable insights grounded in prior experience.
Building on this idea, researchers have developed interactive case-based retrieval systems that surfaced prior instances based on similar features or concepts through explanations.
Applications include clinical diagnosis review \citep{Cai2019-zx, Suresh2022-lv}, auditing dataset annotation quality \citep{Kenny2021-xb}, identifying model vulnerabilities \citep{Sun2023-dm}, and refining chatbot behaviors \citep{Miranda2021-dj}. 

Consider \citet{Cai2019-zx}'s case-based retrieval tool for medical diagnosis. 
Rather than evaluating AI outputs directly, pathologists used the tool to refine search queries that retrieved similar scans with confirmed diagnoses, supporting their interpretation of the current image.
Pathologists improved queries by either cropping salient regions from normative scans or adjusting the feature importance of known medical concepts.
Here, the primary XAI objective is to improve diagnostic accuracy for pathologists (i.e., why AI merits use) rather than understanding and verifying AI predictions (i.e., how AI works). 

\subsection{Credibility} \label{subsec:Ethos} 


Explanations can influence user perceived trust in AI systems when designed to project credibility.
For the first goal of understanding how AI works, we identify three credibility-focused strategies: \textbf{Authority Framing (\ref{sec:authFrame})}, \textbf{Cultural Relatedness (\ref{sec:culturalRel})}, and \textbf{Numerical Association (\ref{sec:NumerAsso})}.
To communicate why an AI system is trustworthy and worth adopting, we identify two additional strategies: \textbf{Technical Assurance (\ref{sec:TechAssur})} and \textbf{Social Endorsement (\ref{sec:socEnd})}. 


\subsubsection{Authority Framing} 
\label{sec:authFrame}
HCI researchers have examined how different communication styles affect user acceptance of machine-generated advice.
These styles include direct versus implicit recommendations \citep{Rau2009-zk} and languages that signal authority or expertise \citep{Metzger2024-nw, Saunderson2021-sl}. 
Naturally, designers have integrated authoritative language and tone into explanations to enhance their perceived credibility  \citep{Cambria2023-af, Okoso2025-ja}.
We refer to this design strategy as \textbf{Authority Framing}.

For example, an AI-recommended insulin dose can be phrased persuasively to make patients more likely to trust and act on the recommendation. 
In a detached or impersonal tone, the technical explanation might read:
\begin{quote}
    ``Because your planned alcohol intake exceeds one unit, the system recommends a lower dose of insulin. A normal dose would have been appropriate if the intake were one unit or less.'' (Example adapted from \citet{Van_der_Waa2021-he})
\end{quote}

Alternatively, the same explanation can be delivered in a more supportive tone, echoing a nursing communication style that balances empathy with expertise.
\begin{quote}
``Planning to enjoy more than one drink? The system suggests making a small adjustment to your insulin dose to keep you feeling your best. If you only have one unit or less, your usual dose is just right. We're here to support you in making healthier choices!''
\end{quote}
 
Another example is an AI tour guide by \citet{Diaz-Rodriguez2020-es}, which combined expressive language with domain expertise to explain the historical context of paintings, sculptures, and other artworks.
Visitors responded more positively when explanations conveyed both factual accuracy and emotional resonance.
Thus, authority framing might be effective for personal contexts, where users interpret and act on system guidance that directly affects their behaviors. In these cases, combining expertise with empathy not only increases the perceived credibility of the recommendations but also makes users feel emotionally supported in making sensitive or value-laden decisions \citep{Okolo2024-et, Komatsu2020-xz}.


\subsubsection{Cultural Relatedness} \label{sec:culturalRel}
Beyond linguistic styles, XAI researchers have also integrated visual elements (e.g., color and shape) into explanations to project a credible tone, voice, or personality. 
Users are more likely to understand and trust explanations when presented visual elements resonate with their beliefs, values, or cultural identities. We refer to this strategy as \textbf{Cultural Relatedness}.

For example, \citet{Okolo2024-et} studied the deployment of a neonatal jaundice\footnote{Neonatal jaundice is a prevalent health condition in newborns, marked by yellowing of the skin and eyes.} detection application in rural Indian clinics.
During a co-design workshop, community health workers raised concerns about the default color scheme used in LIME \citep{Ribeiro2016-dl} and SHAP \citep{Ghorbani2019-ds} visualizations. 
In these explanatory visualizations, red simply denoted a classification outcome.
However, health workers found the red color schemes confusing because red typically signifies severity and danger in medical contexts.
To address this concern, workers applied traditional Hindu colors and styles to replace default color schemes in LIME and SHAP. 


To avoid misinterpretations associated with colors, \citet{Okolo2024-et} experimented with other visual techniques to convey feature importance, such as textual descriptions and different shapes (e.g., squares and circles). Clinicians found that replacing color with text improved clarity, but using shapes caused new confusion, as some interpreted quantity (e.g., number of circles) as a proxy for condition severity. These findings emphasize the need for culturally informed visual designs to avoid unintended miscommunications.



\subsubsection{Numerical Association} \label{sec:NumerAsso}
Another common technique to project credibility is \textbf{numerical association}, where explanations embed numbers, scores, or other metrics to reinforce perceived objectivity and authority. 
Numbers often carry a rhetorical power: they signal precision, rigor, and algorithmic sophistication. \added{A well-known illustration is the phrase ``Lies, damned lies, and statistics,'' which captures how numbers can often be selectively chosen to bolster nearly any desired argument.}\footnote{Lies, damned lies, and statistics: \url{https://en.wikipedia.org/wiki/Lies,_damned_lies,_and_statistics}}

In the context of XAI, numerical association is effective in quantitative contexts (e.g., performance reviews, financial scoring, or risk assessments), where users may benefit from the perceived objectivity of numbers to legitimize, clarify, and compare AI recommendations. On the other hand, users may also mistakenly equate numerical information with higher system competence, even if they do not fully understand its statistical meaning \citep{Li2023-xo}.

For example, \citet{Ehsan2024-lu} implemented three different styles of AI explanations to evaluate how users perceive an AI agent's intelligence. Regardless of participant backgrounds in AI knowledge, users consistently rated the agent with numerical explanations as more intelligent and reliable, even when they lacked a clear understanding of the presented metrics. The authors attributed this \textit{unwarranted faith} to preconceived associations between numerical information and algorithmic thinking processes, reinforcing a perception of objectivity without true interpretability. 


These examples highlight a key challenge: managing the tradeoff between credibility (ethos) and logical clarity (logos). While design strategies like cultural relatedness and numerical association can increase users’ perceived trust in AI systems, they do not always support a better understanding of how the system works. In some cases, as observed in \citet{Okolo2024-et} and \citet{Ehsan2024-lu}, credibility-focused strategies may even obscure how AI systems work by introducing misinterpretations or inducing false confidence. See Section 6.3.1 for further discussion on responsibly signaling authority.



\subsubsection{Technical Assurance} \label{sec:TechAssur}
In practice, AI developers often employ \textbf{technical criteria} to validate predictive accuracy and fairness. 
Examples of technical standards include mitigating model bias \citep{Gupta2023-jx} and ensuring robustness against adversarial inputs \citep{Gupta2024-nq}.
Incorporating these criteria into explanations can foster perceived user trust by clarifying the AI system's legitimacy and justifying why it merits use.
For instance, \citet{Mitchell2019-lf}’s model cards and \citet{Ribeiro2020-vn}’s model behavior checklists exemplify how technical reliability can be assessed and communicated to users. 
Their design strategies convey an implicit message of assurance: \textit{We have made every effort to ensure this AI model is reliable so that you can use it safely and confidently.} 

While primarily for developers, tools like model cards and behavior checklists also serve a broader purpose. They signal that the model aligns with rigorous technical standards, which can subtly influence perceived user trust and adoption in practice. Like car manuals, which are rarely used but still inspire confidence, the real value of these technical artifacts lies in conveying assurance and reliability. Moreover, these artifacts also help brand a company's identity around commitments to quality and responsible innovation.

To support complementary human-AI teaming, designers have incorporated technical criteria alongside adaptive explanations (introduced in Section \ref{subsec:Logos}).
For example, \citet{Zhang2020-pf} incorporated AI confidence scores alongside feature-based explanations to examine whether displaying confidence levels would increase user likelihood of accepting AI predictions. 
Similarly, \citet{Bansal2020-zw} adapted the presentation of explanations based on the system's confidence, showing explanations only when the model's predicted confidence was high. 
Both studies found that participants heavily relied on these technical signals to guide their acceptance of AI predictions. 

However, when exposed to technical indicators such as confidence scores, users may blindly follow AI predictions without critically assessing explanations. 
Despite this concern, researchers have yet to identify the most effective strategy to mitigate over-reliance. Some researchers have explored adding cognitive forcing functions to encourage more critical engagements with AI explanations \citep{Gajos2022-ns}. 
Other researchers attribute the functional utility of explanations to domain experts, suggesting that their effectiveness depends on the user prior knowledge \citep{Bansal2020-zw, Wang2021-hv}. 
However, these works \added{might omit a further inquiry when using credible signals to achieve complementary performance:} \deleted{highlight a key trade-off between AI explanations' functional utility and persuasive credibility} \textit{If users ultimately learn to perform the task independently, what other complementary value does the AI system continue to provide?} 

We argue that engineers and technical developers should be responsible for ensuring that technical criteria support the responsible use of AI. Given the limited expertise of lay users, it is unrealistic to expect them to critically evaluate the reliability of technical signals.
Once technical signals can faithfully convey the AI's reliability, they should be carefully integrated into user-facing designs to promote assurance and confident use.

\subsubsection{Social Endorsement} 
\label{sec:socEnd}
Beyond incorporating technical assurances, researchers have also employed \textbf{social endorsements} to strengthen user perceived trust in AI systems. 
In organizational settings, social affirmations may exert a greater influence than the explanations themselves on whether users accept or reject AI decisions.

For example, \citet{Ehsan2021-id} developed an explainable AI interface for sales recommendations. Alongside feature-based explanations, the system provided a historical trajectory that showed how past sales teammates responded to the AI recommendation and the rationales behind their choices. 
Participants reported that these design elements enhanced their sense of social validation, transitive trust, and the temporal relevance of AI recommendations within the organizational setting.

While \citet{Ehsan2021-id} framed their design as promoting sociotechnical transparency, we interpret their approach as an example of social endorsement. 
Similar to how product reviews influence consumer behavior \citep{talib2017social}, their design leverages peer adoption to foster trust. 
When new users see that trusted colleagues have adopted and endorsed the system, they are more inclined to perceive the AI as credible and worth using.
Social endorsements can also signal alignment with institutional norms, allowing trust to propagate socially through transitive peer relationships and collaborative culture.



\subsection{Emotional Resonance} \label{subsec:Pathos}


Emotion has historically played a crucial role in design \citep{Ho2012-ng,Norman2007-qr}, serving essential communicative and experiential functions. 
In XAI, researchers have incorporated affective elements into explanations to make them more engaging. 
%
We identify three design strategies to support explaining how AI works: \textbf{Explanation Affect (\ref{sec:exaAffect})}, \textbf{Emotional Cues (\ref{sec:emoCue})}, and \textbf{Protective Framing (\ref{sec:protFram})}. 
Additionally, we identify two strategies to explain why AI applications are emotionally compelling to use: \textbf{Expectation Management (\ref{sec:expMan})} and \textbf{Expressive Interaction (\ref{sec:exprInte})}. 


\subsubsection{Explanation Affect} \label{sec:exaAffect}
\added{
Rather than treating surprise as a purely reactive response to abnormal events, cognitive research suggests that perceived surprise reflects a metacognitive estimate of the cognitive effort required to explain how an unexpected outcome occurred \citep{Foster2015-zg}.
Thus, users' affective responses may differ based on the explanation's perceived completeness, relevance, and cognitive load.
In XAI, prior studies have revealed that} different types of AI explanations (see Appendix \ref{appendix:types of AI explanations}) can inherently evoke varying emotional responses depending on their relevance to the task, a phenomenon known as \textbf{explanation affect}. For example, \citet{Bernardo2023-xv} implemented example-based, feature-based, and rule-based explanations for animal identification.
Participants judged the emotional appeal of each explanation type, highlighting differences in how each form engaged users.
The study reported that example-based explanations elicited stronger feelings of surprise, trust, and perceived usefulness. In contrast, feature-based and rule-based explanations led to reduced emotional engagement.

However, the authors cautioned that these findings might not generalize to other tasks.  
In their study, the red color used in feature-based explanations to mark the Tarsier’s eyes and ears might have unintentionally evoked negative emotions such as fear or discomfort, an effect absent in the presented example-based explanations.
Additionally, the author emphasized that user prior emotional states and attitudes toward AI could strongly influence how different explanation formats affect decision-making and trust. 

\deleted{Thus, in practice, explanation affect emphasizes matching the \textit{emotional fit} between the explanation form and the underlying human concern. In public-facing applications, such as wildlife conservation or sustainability, example-based explanations may be particularly effective, as they evoke empathy and foster personal connections by referencing real-world past cases. In contrast, in more high-stakes settings, such as medical treatment \citep{Van_der_Waa2021-he}, recidivism screening and financial decision-making \citep{Wang2022-rn}, rule-based explanations are preferable due to their capacity to convey structure, fairness, and procedural consistency.}

\subsubsection{Emotional Cues} 
\label{sec:emoCue}
Unlike \textit{explanation affect}, where different types of explanations inherently evoke diverse emotions, some XAI researchers have directly integrated \textbf{emotional cues} into explanations to enhance user engagement. 
Examples of emotional cues include emojis, color, and tone of language.

For instance, \citet{Freire2023-gh} incorporated emojis into AI explanations to increase user engagement. 
\citet{Bernardo2023-xv} compared robotic versus human-like language styles and found that robotic phrasing heightened anxiety and reduced perceived usefulness, while human-like language eased fear and improved perceived utility. 
In some cases, however, emotional responses are triggered unintentionally. For example, \citet{Okolo2024-et} found that the red color used in LIME and SHAP visualizations evoked negative emotions in a medical context, where red is commonly associated with danger, even though this was not the intended message of the explanation.

Additionally, researchers have incorporated emotional cues into interactive AI explanations. For example, in \citet{Wang2023-ar}'s \textit{GamCoach} (introduced in Section \ref{sec:expRec}), the design leveraged different colors, emojis, and visual cues to recommend interactions for algorithmic recourse. Participants found the tool relatively easy and enjoyable to use.

\deleted{As emotional cues draw on visual, linguistic, or stylistic elements to elicit affective engagement, these characteristics can be particularly effective in educational AI tools. When integrated with adaptive explanations, as shown in the augmentation-based writing systems developed by \mbox{\citet{Wambsganss2020-fq}}, emotional cues can enhance memory retention and spark curiosity, thereby enriching students’ learning experiences.}


\subsubsection{Protective Framing} \label{sec:protFram}
Emotional cues are typically designed to elicit positive reactions toward AI explanations, making them more engaging, approachable, or enjoyable. In contrast, \textbf{protective framing}\footnote{We adopt this term from \citet{Fokkinga2012-vq}, who explored how people use negative emotions in everyday life to achieve goals} relies on negative emotions like skepticism or caution to encourage more critical reflection of AI explanations.
Our narrative literature review did not identify design examples that explicitly employed this strategy. 
Instead, many studies introduced algorithmic aversions or AI system failures as independent variables to examine user acceptance of AI decisions and engagement with explanations. \added{Findings from these studies suggest that protective framing could be used to prompt users to critically evaluate explanations.}
\deleted{Thus, we propose protective framing as a possible direction for future XAI design.}


For example, \citet{Ebermann2023-hx} and \citet{Ben_David2021-kn} found that negative feelings such as disappointment, annoyance, or frustration often arise when users encounter unexpected AI outcomes or failures. 
These adverse emotions can heighten user attention to explanations. 
Notably, \citet{Ben_David2021-kn} observed that following an AI failure, participants were significantly less likely to accept subsequent AI advice when explanations were shown, compared to when no explanations were shown. This suggests that explanations can amplify critical scrutiny under negative affect. 

\deleted{Building on this observation, we could imagine a design scenario illustrating how protective framing could facilitate more critical user engagement. When the AI prediction contrasts with a user's expectations, a system alert might state:} 

\deleted{``The AI has predicted a different outcome, but it's important to remember that both humans and AI can sometimes make mistakes. We encourage you to review the detailed explanation behind the AI's decision...[show AI explanations]''} 

Additionally, even when the AI prediction aligns with the user's decision, the system can still deliver a cautionary alert, reminding users of the potential for AI to make errors and encouraging them to review the explanation critically.
This cautionary disclaimer has been shown to act as a cognitive forcing function, effectively slowing down the decision-making process and prompting users to process explanations more carefully \citep{Gajos2022-ns, Bucinca2021-uc}.

For practitioners, protective framing helps directly safeguard against overreliance on AI systems.
Specifically, failures could serve as a protective frame, prompting users to explore AI explanations proactively before finalizing their decisions. 
This idea reflects an alternate interpretation of the cost-benefit model described by \citet{Vasconcelos2023-cm}. 
While their work focused on reducing the cost or increasing the benefits of verifying AI predictions (through explanations), triggering negative emotions emphasizes the consequences of neglecting verification, effectively increasing the perceived benefit of careful review. 

\subsubsection{Expectation Management} \label{sec:expMan}
In real-world AI applications, explanations are commonly used to\textbf{ manage user expectations}. 
Before interaction, explanations can help justify why the AI system is worth using by outlining its strengths and limitations under intended use cases. After interaction, explanations clarify how the AI works by outlining the factors influencing its recommendations.
This mirrors how people use explanations to set and adjust expectations, maintain shared understanding, and resolve ambiguity in everyday communication. 


For example, \citet{Kocielnik2019-sq} found that presenting AI explanations upfront could lower user expectations of AI performance, thereby reducing dissatisfaction during AI breakdowns. 
However, this emotional remedy may not be effective when explanations are not perceived as practically helpful. For example, \citet{Ebermann2023-hx} found that participant mood scores significantly declined when explanations justified alternative decisions rather than the AI's predictions, creating cognitive dissonance and undermining user satisfaction. 
Interestingly, despite the observed negative emotional impact, participant evaluations of system usefulness (i.e., self-rating on the question ``to what extent was my [AI] support helpful'') actually improved.
This divergence suggests that emotional discomfort does not always indicate poor system design. Like how students may resist challenging coursework yet still benefit from it, friction in human-AI interaction may play a productive role by encouraging more critical system evaluations.

For practitioners, expectation management can calibrate user expectations, preempt dissatisfaction, and maintain trust when introducing a novel AI system under early development (i.e., when system performance may be unstable). Additionally, expectation management can help mitigate algorithmic aversion \citep{Hou2021-kp, Cheng2023AA} by clarifying an AI system’s strengths, limitations, and appropriate boundaries of use, making the system appear more transparent and aligned with user concerns about overreach or misuse.

\subsubsection{Expressive Interaction} \label{sec:exprInte}
Expectation management primarily treats explanations as affective band-aids to remediate AI breakdowns. 
While this reactive use of explanations can be effective, designing with emotional elements also calls for a more proactive and holistic perspective. 
Beyond remediation, AI explanations can enable \textbf{expressive interaction}, a design approach that sustains emotional engagement throughout the user’s interactive journey with the AI system.

For example, \citet{Alvarado2022-ys} developed a humanoid controller to interact with a movie recommendation algorithm. Users can accept or reject AI-suggested movies through various physical controls, such as pressing, holding, or shaking the controller. Notably, a long press at the controller's center triggers the system to explain why a particular movie was recommended based on the user's viewing history or personal preferences. 
This gesture provides a subtle way to address user curiosity about why a particular recommendation was made.
More importantly, AI explanations are woven into a larger embodied interaction, functioning as a natural part of the experience rather than as isolated add-ons. 

Expressive interaction treats emotion not as a remedy, but as a generative force that actively shapes how users engage with AI systems. Rather than smoothing over frictions, this design strategy aims to seamlessly integrate emotional responses as a natural part of human-AI interplay. Through repeated interactions, users develop personal rhythms with the system: deciding when to seek explanations, how much to rely on them, and what emotions to attach to different types of feedback. In this way, explanations become part of an evolving emotional narrative, rather than static tools for understanding model behavior. As a result, expressive interaction naturally supports long-term engagement applications by enabling personalized explanation experiences that adapt to evolving user needs, emotions, and goals.



\section{Discussion} \label{sec:Discussion}

By synthesizing design strategies employed in prior XAI systems, we have identified a pluralistic design landscape, reflecting how researchers have explained how AI works and why AI merits use. 
Since the reviewed literature captured only a subset of existing design strategies, our synthesis should be viewed as a starting point rather than a comprehensive catalog.
As XAI evolves, new strategies will likely emerge for different rhetorical appeals. 
In this section, we discuss the contributions of these diverse strategies to existing XAI research (Section \ref{subsec:task relevance} and \ref{subsec:artistic practice}), the caveats of identified design strategies, and limitations of our review process (Section \ref{sec:limitations}).

\subsection{Centering XAI Around Real User Intent and Task Relevance} \label{subsec:task relevance}

As described in our rhetorical framework (Section \ref{sec:Framework}), we extend the traditional goals of XAI beyond merely clarifying model behavior to include another dimension: \textit{explaining why AI merits use}. Note that this goal is better understood as a design objective rather than a technical optimization. 
When explaining \textit{why} AI merits use, it is important to consider when AI is and is not appropriate, and its degree of suitability for a given task. Factors such as model reliability \citep{Petch2022-ju}, error consequences \citep{Ehsan2022-fz}, and alignment with user goals help users make informed adoption decisions.
As highlighted in Section \ref{sec:Results}, prior researchers have already employed various design strategies to support this objective. Some used explanations as recommenders or retrievers, offering actionable insights for real-world problem solving. Others applied credible or emotional framing to make AI applications more compelling and persuasive. 
\added{
That said, \citet{Wardatzky2025-xo} also caution that since many existing XAI studies rely on crowdsourced participants for evaluations, their findings may not generalize to actual users. 
}

A natural direction for future XAI research is to incorporate real user intent, task relevance, and model reliability more explicitly as design criteria when selecting and evaluating explanations. In particular, researchers might consider: 1) whether a user's strong motivation to adopt AI for a given task creates a meaningful need for explanations, 2) to what extent that need varies by task contexts, and 3) how reliably the AI can be used in conjunction with explanations that promote safe and informed decisions. Connecting user intent, AI robustness, and explanatory relevance in this way can enhance both the interpretive value and empirical rigor of XAI research.

For example, when examining the effects of different AI explanations on sentiment classification, recidivism prediction, or other tasks, the first set of questions could start from: \textit{Who are the targeted users of these AI tasks, and why would they want AI to assist them?} 
Answers to these questions could provide specific use cases to further investigate users' explanatory needs: \textit{Why would they care about understanding individual predictions or model global behavior in these tasks? What AI explanations and design strategies could better achieve their goals for using AI?} 
Even when traditional AI explanations fall short, answering these questions can reveal new user needs and prompt designers to rethink the problem space, leading to more relevant solutions \citep{Zimmerman2022-xy}.
This shift in perspective opens opportunities to repurpose XAI techniques to support emerging AI applications beyond their original intent \citep{Liu2023-ur}.


\subsection{Perceiving Explanation Design as a Reflective Rhetorical Practice} \label{subsec:artistic practice}

In HCI, there is growing interest in promoting stakeholder participation and integrating cross-disciplinary knowledge when developing XAI systems \citep{Malizia2023-fz, Mohseni2021-bw, Dhanorkar2021-ah}. Within these collaborative settings, \added{explanation design becomes less about technical optimizations and more about navigating diverse, and sometimes competing, stakeholder goals. Our framework addresses this challenge by synthesizing how rhetorical strategies have been operationalized in prior XAI work. We also document both positive and problematic user effects that rhetorical strategies may produce. Rather than prescribing what constitutes best practice, we offer a conceptual lens for critically examining explanation designs across contexts. 
}

\added{From this perspective,} designers can \added{draw}\deleted{play a critical role in addressing diverse stakeholder needs by drawing} on the strategies synthesized in Section \ref{sec:Results} \added{to balance different rhetorical appeals in response to situated needs}. \deleted{Specifically, designers can strive to balance different rhetorical appeals to accommodate the diversity and flexibility of human needs.} For example, in some use cases, logical comprehension of the AI system and its practical utility take precedence, with credibility and emotional factors serving a secondary role. This is often the case when designing AI-powered analytical tools for technical users, such as developers or data scientists. In contrast, for consumer-facing AI applications, credibility and emotional resonance may be more influential in enhancing user engagement and trust. 

Additionally, different design strategies can help mitigate each other's  limitations. 
For example,  \citet{Gaole2025-wo} observed that users tended to overrely on the AI system when presented with conversational or mechanistic explanations, which created an illusion of explanatory depth. To address this, designers may incorporate emotional strategies such as protective framing or expectation management to encourage more critical user engagement. However, integrating and balancing multiple design strategies to meet diverse stakeholder goals remains a complex, evolving challenge for future work. For example, an XAI design solution that worked well at one point, or for a particular stakeholder group in a specific domain, may become inadequate as the application context evolves, necessitating new design responses \citep{Friedman2008-yk, Zimmerman2022-xy}. \citet{Buchanan1992-bo} argued that design is a form of inquiry, where the actual problems emerge through continuous user engagement, reflection, and iterative assessment. This perspective is particularly relevant for designing AI explanations, which increasingly involves collaborative and sociotechnical considerations (discussed in Sections \ref{subsec:stakeholder-centered XAI} and \ref{subsec:sociotechnical XAI}). Thus, addressing this challenge may require designers to use an iterative, exploratory design process (e.g., research through design \citep{Zimmerman2014-es}) combined with creative strategies to navigate potentially competing and shifting needs. 


\subsection{Misuses of Rhetorical Strategies} 
\label{sec:Misuses}

While rhetorical strategies can make AI explanations more compelling, they inadvertently carry ethical risks. If misapplied without guardrails, these strategies can blur the line between persuasion and manipulation, raising concerns about transparency, fairness, and user autonomy. Several risks were discussed earlier when we introduced each rhetorical strategy \added{(e.g., shallow or distorted understanding from oversimplified explanations in Section \ref{sec:isoPre}, the illusion of explanatory depth from adaptive and mechanistic presentation in Section \ref{sec:adapPre}, and mismatches between color associations and cultural contexts presented in Section \ref{sec:culturalRel})}. In this section, we briefly outline some broad patterns of potential misuse across the three rhetorical dimensions (logos, ethos, and pathos). We do not expect to provide a comprehensive typology, but only a starting point for caution and reflection. As rhetorical XAI becomes more widely adopted, we recommend a deeper analysis of how to balance persuasive design with responsible practice as future work.


\subsubsection{Logos} One risk pattern in logos is the use of selective logical appeals to create \textit{explanatory fallacies}, where explanations appear logically coherent but are actually incomplete or misleading. These fallacies may arise from intentional disinformation or unintentional misinformation conveyed in the explanations. For example, \textit{explanations as recommenders} (\ref{sec:expRec}) may only present advantageous factors that support adoption while omitting adverse side effects. Consider a clinical decision-support system that might explain its recommendation for a new medication by highlighting improved recovery rates, while failing to mention common side effects such as nausea or fatigue. As a result, explanatory fallacies may conceal relevant model reasoning and downplay potential risks, undermining interpretive fidelity \citep{Morrison2023-to}. Additionally, the decision-support system itself could accidentally omit vital information or hallucinate false information due to limitations in its training data. Thus, it is important for clinicians to be actively involved in this process to assess whether these explanations are accurate, appropriate, and sufficient \citep{Amann2020-kn}.


\subsubsection{Ethos} Credibility appeals may be deceptive when the source of trust is unverified, unearned or a previously trusted authority loses trust. Strategies like \textit{authority framing} (\ref{sec:authFrame}) or \textit{social endorsement} (\ref{sec:socEnd}) may leverage institutional branding, peer behavior, or delayed fact-checking to signal legitimacy, regardless of actual system quality. For instance, a university admissions algorithm might present its decisions with a badge like ``Certified by Admissions Experts,'' even if such certification has not undergone verification. By applying the language and visual cues of credibility, unethical AI systems may foster misplaced trust by masking the absence of verified authority.


\subsubsection{Pathos} For pathos, \textit{emotional cues} (\ref{sec:emoCue}) risk manipulating user behavior when they reinforce decisions that may not align with user goals or best interests. For instance, a personal finance assistant may use cheerful animations or a green thumbs-up icon to reward users for complying with its savings advice. While such cues foster positive affect, they may also encourage users to accept AI recommendations uncritically, reinforcing confirmation bias  \citep{Mercier2011-jf, El-Assady2022-xa}. This illustrates how affective nudges can substitute emotional reassurance for critical thinking, deceptively fostering unwarranted trust without genuine understanding.


\vspace{0.2cm}
\noindent 
These risks of manipulation recall the analogy of designers as rhetoricians, a role increasingly associated with negative connotations in marketing \citep{Miles2018-ay}, adversarial propaganda campaigns \citep{Farrell2018-yx}, and foreign influence operations \citep{Starbird2019-yp}.  
%
%
As with other aspects of AI, potential benefits often come with potential risks. We believe it important to both acknowledge such potential risks and 
emphasize the ethical responsibilities designers should uphold when employing these strategies to support stakeholder goals. Following Buchanan \citep{Buchanan1985-qy, Buchanan1985-zf}, we position rhetoric as a means to support reasoned invention and explore meaningful design alternatives. At the same time, we recognize that any persuasive strategy carries the risk of misuse. Introducing rhetorical approaches in explanation designs obliges us and future designers to take responsibility for anticipating how such methods could intentionally or inadvertently mislead or deceive. Future research should not only explore how rhetorical design strategies could be balanced with ethical concerns, but also critically examine safeguards, accountability mechanism, and the contexts where these strategies are most appropriately applied. One approach could involve framing them as dark patterns \citep{Gray2018-bt, krandarkbench2025} to warn future XAI designers.



In sum, by exploring the potential of rhetoric in explanation design, we argue that design strategies are not inherently benign or deceptive; rather, their ethical impact largely depends on how they are employed. 
%
\added{As Melvin Kranzberg famously stated, ``Technology is neither good nor bad; nor is it neutral'' \cite{kranzberg1986technology}.}  
When grounded in human-centered values and established principles of responsible communication, such as sustainability \citep{Blevis2007-cr}, dignity \citep{Kim2021-jl, Buchanan2001-br}, and social inclusiveness \citep{Costanza-Chock2020-rl, Thorpe2011-vm}, rhetorical explanations can support ethical engagement with AI systems and mitigate risks of misuse. This perspective frames rhetorical strategies as communicative tools that can be informed by the wider canon of communication practices, highlighting the responsibility of designers in shaping their applications.

\subsection{Study Limitations}
\label{sec:limitations}

\subsubsection{Narrative Review Process}
Future work can refine our literature review process as rhetorical design gains broader recognition in the XAI community and new design strategies are proposed and systematically evaluated.

In our review, we categorized papers according to two primary XAI goals: (1) explaining how AI works, and (2) explaining why AI merits use. We observed that many studies targeting the first goal also included measurements relevant to the second, such as perceived AI usefulness and user trust in the AI system. For example, technical assurance was frequently incorporated in XAI studies to measure how users perceived different explanations, even though these studies did not explicitly articulate user intent and the task relevance (e.g., \textit{why participants might use AI to complete the task}). Nonetheless, our axial coding incorporated both quantitative and qualitative findings from prior studies to understand the influence of different design strategies in promoting the AI's benefits. Future reviews might consider more stringent inclusion criteria to ensure more precise alignment between study design, user intent, and task relevance.

Additionally, as described in our axial coding process, we mapped user study measurements from prior work to the three rhetorical appeals in our framework. However, this mapping process was inherently conceptual and subjective, incorporating author biases. On the other hand, this mapping also suggests an opportunity to develop a more comprehensive set of measurement tools explicitly aligned with the dimensions of logical reasoning, credibility, and emotional resonance. Future research could systematically examine how existing evaluation metrics align with each rhetorical appeal, followed by a scoping review to understand the effects of different design strategies better.

\subsubsection{Practical Implementation Gaps}
\label{sec:implementationGaps}


In this work, we have introduced a new taxonomy to organize and reflect on rhetorical strategies in XAI, encouraging researchers to consider designing explanations to not only explain how AI works but also justify why AI merits use. For each strategy, we highlighted potential applications and settings where these approaches may be beneficial. However, determining which strateg(ies) to apply in a given scenario remains an open challenge. Triage and selection will likely depend on context-specific factors and we leave this important direction for future work.

\section{Conclusion} \label{sec:Conclusion}

We have proposed conceptualizing AI explanations as rhetorical artifacts, designed not only to support user understanding of AI systems but also to promote appropriate AI adoption as another dimension of responsible AI. Drawing on the concept of rhetorical design, we developed a rhetorical framework that guided our coding process for a narrative literature review. This review synthesized a range of design strategies used in prior XAI studies to make explanations logically convincing, credibly appealing, and emotionally engaging. By critically examining the strengths and limitations of these strategies, our work extends the design space of XAI by providing designers with new tools to address the complex needs of diverse stakeholders and the broader sociotechnical challenges involved in building XAI systems.


\section*{Acknowledgments}

We thank Ahmer Arif, Jacek Gwizdka, James Howison, Yujin Choi, and Yiwei Wu for their valuable comments. This research was supported in part by Good Systems\footnote{\url{http://goodsystems.utexas.edu/}}, a UT Austin Grand Challenge to develop responsible AI technologies. The statements made herein are solely the opinions of the authors and do not reflect the views of the sponsors.

\bibliographystyle{ACM-Reference-Format}
\bibliography{reference}


\appendix

\section{Additional Statistics, Tables, and Explanations of the Narrative Review Process} \label{appendix:review}

\added{Table \ref{tab:database_articles} summarizes the number of articles collected across our literature search. Because this study follows a narrative review approach, we did not exhaustively screen or read all retrieved papers. Instead, our review proceeded iteratively over time, with cycles of reading and coding as our understanding of the field deepened.}

\subsection{Additional Explanations for Phase 1 Screening Conceptual and Technical Literature} \label{appendix:phase 1}
\added{In the first round of screening and close reading, we focused on distilling core technical concepts, design paradigms, and conceptual frameworks in the XAI literature. Table \ref{tab:categorization of XAI survey and framework} lists the design-oriented survey and framework papers examined during this initial phase. The resulting categories reflect a bottom-up thematic grouping of how prior work frames XAI surveys and frameworks. We synthesize insights from this phase in Section \ref{sec:XAI design}. At the same time, we also examined survey and framework papers from a more technical perspective. This reading helped us compare taxonomies and emphases across the AI and HCI communities. We synthesize these technical perspectives in Appendix \ref{appendix:XAI terminologies} and Appendix \ref{appendix:types of AI explanations}.}

\begin{table}[ht]
\centering
\begin{tabular}{@{}l r@{}}
\toprule
\textbf{Database} & \textbf{Number of Articles} \\
\midrule
Web of Science Core Collection & 92 \\
ACM Digital Library & 88 \\
ScienceDirect & 84 \\
Taylor \& Francis & 9 \\
\midrule
\multicolumn{1}{@{}l}{Total} & 273 \\
\multicolumn{1}{@{}l}{Publication date} & (1987 to 2025) \\
\bottomrule
\end{tabular}
\vspace{0.2cm}
\caption{Articles were identified through electronic database searches conducted in two phases. Paper title or abstract that did not have ``Explainable AI'' OR ``interpretable AI'' AND ``Design'' were excluded. In total, 273 unique papers were included.}
\label{tab:database_articles}
\end{table}

\begin{table}[ht]
\centering
\begin{tabular}{@{}p{0.22\linewidth} p{0.70\linewidth}@{}}
\toprule
\textbf{Category (\textit{N = 29})} & \textbf{Representative Papers} \\
\midrule
User-centered (\textit{15}) & 
\textit{Understanding user needs:} \newline
\citet{Sanneman2020-hb}, \citet{Liao2020-da}, \citet{Xin_He2023-dj}, \citet{Brennen2020-kw}, \citet{Mucha2021-xb}, \citet{Hagras2018-hx}, \citet{Vera_Liao2021-zi}, \citet{Kaplan2024-td}, \citet{Vera_Liao2022-ro}  \newline
\textit{Alignment between human cognition and model understanding:} \newline
\citet{Wang2019-we}, \citet{Miller2023-nc}, \citet{Singh2021-ws}, \citet{Miller2019-zh}, \citet{Bertrand2022-xg}, \citet{Dragoni2020-tn} 
\\

Stakeholder-centered (\textit{5}) &
\citet{Dhanorkar2021-ah}, \citet{Hoffman2021-ek}, \citet{Mohseni2021-bw}, \citet{Preece2018-ql}, \citet{Tomsett2018-ij}
\\

Socio-technical (\textit{5}) &
\citet{Sokol2020-lo}, \citet{Ehsan2021-id}, \citet{Ehsan2020-gz}, \citet{Ehsan2023-pu}, \citet{Brandao2019-qi}
\\

Others (\textit{4}) &
\textit{Forms of XAI interactions:} \newline
\citet{Bertrand2023-xy}, \citet{Chromik2021-ef}, \citet{Xu2023-qd} \newline
\textit{Goal-oriented framework:} \newline
\citet{Suresh2021-qq}
\\
\bottomrule
\end{tabular}
\vspace{0.2cm}
\caption{Categorization of XAI Survey and Framework Papers}
\label{tab:categorization of XAI survey and framework}
\end{table}

\subsection{Additional Explanations for Phase 2 Screening Empirical Literature and Mapping Rhetorical Effects} \label{appendix:phase 2}
\added{In the second round of screening and close reading, we shifted our focus to design-oriented and empirically driven XAI studies. During this phase, we iteratively coded each paper using our proposed framework (Section \ref{sec:Framework}). Through a combination of deductive and inductive coding, we identified a subset of papers that satisfy one or both of the explanatory goals defined by our framework. These papers were therefore included in the annotated collection presented in Appendix~\ref{appendix:codebook}. Below, we outline several types of exclusions, such as instances where there is insufficient evidence to substantiate rhetorical strategies and their persuasive impact, as well as design interventions that are not rhetorical in nature.}

\subsubsection{Ambiguity in Design-Effect Associations}
\added{Although some studies highlight potential rhetorical effects, the associations between XAI designs and these effects are unclear due to differing study emphases, such as reporting overall challenges of XAI tools \citep{Kaur2020-ds}, focusing on the XAI design process \citep{Bhattacharya2024-qb}, or synthesizing end-user design suggestions \citep{Wang2025-xy}. For these studies, we cannot confidently claim that specific user effects are direct outcomes of particular XAI designs.} 

\subsubsection{Methodological Interventions vs. Rhetorical Strategies}
\added{Second, some studies employed strategies that differ from the underlying rhetorical assumptions. This line of research often uses methodological interventions that force rationality beyond natural persuasion, including play \citep{Morrison2023-uh}, cognitive forcing \citep{Gajos2022-ns, Bucinca2021-uc}, or other incentives like financial rewards \citep{Vasconcelos2023-cm, Nguyen2018-az}. In contrast to rhetoric that acknowledges irrationalities (e.g., bounded rationality) as inherent to real-world AI use, these methodological interventions reduce human heuristics by compelling users toward deliberate, analytical processing. Note that some prior XAI studies integrate both approaches (methodological interventions vs. rhetorical strategies) and evaluate user effects of AI explanations in different studies, such as \citep{Vasconcelos2023-cm, Nguyen2018-az} included in our codebook. Methodological interventions provide valuable insights as a complementary approach of rhetorical strategies to influence appropriate user evaluation of XAI. We incorporated them when describing the rhetorical strategies in our analysis.}

\subsubsection{Thematic Saturation}
\added{We stopped further coding once newly examined papers no longer yielded substantively new rhetorical strategies, indicating thematic saturation. This includes several papers that, during the screening process, were found to align with existing design strategies, such as \citep{Kim2023-ds} for different user feedback on isolated explanations, \citep{Radensky2022-kz, Wang2022-hx, Smith-Renner2020-mc} for explanations as recommenders, and \citep{Hadash2022-xo} for framing effects. Because this round was not exhaustive, we acknowledge that some omitted papers may still represent promising candidates for identifying new rhetorical strategies that could complement the existing set summarized in Table \ref{tab:rhetorical_strategies}.}

\section{XAI Technical Concepts and Taxonomies} \label{appendix:XAI terminologies}

Both AI and HCI researchers have acknowledged that making AI systems interpretable to people can foster more trustworthy and accountable use of AI \citep{Markus2021-fz, Molnar2020-uo, Jacovi2021-id, Busuioc2021-gp}. \citet{Doran2017-gd} categorize AI systems into three types --- \textit{opaque}, \textit{interpretable}, and \textit{comprehensible} --- based on the level of transparency they offer in explaining automated decisions. In this taxonomy, interpretability refers to how well users can understand the internal processes by which an AI system transforms inputs into outputs. Other taxonomies of interpretability frame this input-output relationship through alternative lenses such as explanation scope, algorithmic techniques, or interactive format. 

A common taxonomy based on the explanation scope is whether AI explanations are \textit{local} or \textit{global} \citep{Molnar2020-uo, Mohseni2021-bw, Dwivedi2023-oc}. \textit{Local} explanations explain why a specific input maps to an observed output. In contrast, \textit{global} explanations provide an overarching view of how the AI generally maps inputs to outputs. Orthogonally, scope can also vary by model generality. Another foundational taxonomy differentiates \textit{model-specific} techniques, which are tailored to a particular model, from \textit{model-agnostic} techniques, which can be applied to any model. 

Different XAI techniques lead to diverse types of explanations, which HCI researchers have evaluated with respect to their effects on users. These explanation types include \textit{example-based}, \textit{feature-based}, \textit{counterfactual}, \textit{rule-based}, and \textit{concept-based} explanations (see \textbf{Appendix \ref{appendix:types of AI explanations}} for further details). Each supports a distinct form of logical reasoning, reflecting the underlying algorithmic mechanisms that generate the explanation \citep{Wang2019-we}. 

Given the wide range of human-AI interactions enabled by explanations \citep{Bertrand2023-xy, Xu2023-qd, Chromik2021-ef}, many user studies adopt two primary forms of explanations --- \textit{static} vs. \textit{interactive} --- to examine how different explanation types influence user acceptance or foster appropriate trust in individual AI predictions. Static explanations typically present fixed outputs for given inputs and allow participants to engage with them over a limited number of trials. In contrast, interactive explanations enable users to actively engage with the system by adjusting inputs or model parameters, allowing them to explore alternative scenarios and gain a deeper understanding of the model’s behavior \citep{Liu2021-ds, Bertrand2023-xy, Chromik2021-ef, Nguyen2018-az}.

While there is debate about the necessity of AI explanations, many researchers view XAI as a promising pathway for promoting safe, trustworthy, and accountable use of AI. For instance, when humans collaborate with AI to perform different tasks, explanations can help users understand model behavior, enabling them to judge when to accept or reject AI recommendations. This, in turn, fosters appropriate user trust and supports complementary human-AI teaming  \citep{Lee2004-kt, Yang2020-wm, Wang2021-hv, Schemmer2023-zy}. To support cooperation, HCI researchers typically draw on human-centered principles when designing XAI systems (HCXAI). HCXAI emphasizes identifying user needs and contexts of use to inform whether, when, and how different explanations should be presented \citep{Ehsan2020-gz, Ehsan2023-HCXAI, Ehsan2024-HCXAI}. 

\section{Types of AI Explanations} \label{appendix:types of AI explanations}

We briefly summarize AI explanations designed in a static form, including example-based, feature-based, counterfactual, rule-based, and concept-based explanations. These explanations serve as the basic design unit for more complex and interactive AI explanations \citep{Bertrand2023-xy}. Note that this is a design taxonomy based on a logical understanding of how AI works, which might not closely align with the taxonomy used by AI developers, who might prefer the classification of XAI based on interpretable techniques (see Appendix \ref{appendix:XAI terminologies}). 

\subsection{Example-based Explanations} 
Example-based explanations generally refer to the XAI methods that use specific instances that are either retrieved from the training dataset or generated as new data points to explain how AI makes a particular decision. As \citet{Molnar2020-uo} and \citet{Wang2019-we} emphasize, the benefit of these explanations lies in how humans often use examples or analogies to aid in decision-making.  


Researchers have explored different kinds of examples to explain how AI works. The most commonly selected examples are those that share similar features to the input. We refer to these examples as \textit{nearest-neighbors examples} because ``nearest neighbors'' are both intuitive and widely used as a common technique for researchers to retrieve such examples. The classes of these \textit{nearest-neighbors examples} can vary to the predicted output class depending on whether nearest neighbors are directly used in the model influence function. For example, if the prediction model is a k-nearest neighbors model, the examples should share the same output class as the input \citep{Papernot2018-mg, Su2023-ck}. However, if other prediction models are used and nearest neighbor is only used as a retrieval technique, the output class of these examples might differ \citep{Van_der_Waa2021-he, Cai2019-lb, Yang2020-wm, Wang2021-hv, Chen2023-vc}. Therefore, it's important to distinguish these two use scenarios to communicate how AI works. Using nearest neighbors as a retriever can only be interpreted as using examples to examine AI predictions. It might be incorrect to claim that these examples directly influence AI predictions.

\citet{Cai2019-lb} describe another two kinds of examples based on the class difference of nearest-neighbors examples. They are \textit{comparative examples}, examples similar to the input but with alternative output classes, and \textit{normative examples}, prototypical examples with the same class to the input. Both yield a different reasoning process for humans to map AI inputs to outputs based on examples. The idea of comparative examples is similar to counterfactual explanations, informing us how an instance could be changed to alter the AI prediction \citep{Molnar2020-uo, Verma2024-qb} (We elaborate this in Section \ref{sec:CounterExpl}, counterfactual explanations). In contrast, a normative example serves as a benchmark, allowing people to compare the input with a prototypical example to understand how their similarities result in the same output class.

\subsection{Feature-based Explanations} 
\label{sec:FeatureExpl}
Feature-based explanations generally refer to showing feature attribution or influence by indicating which input feature is important to the model global behavior or how it has direct positive or negative influence towards different output classes \citep{Molnar2020-uo, Wang2019-we, Mohseni2021-bw}. Unlike example-based explanations that are mostly local interpretations (i.e., how a specific input affects the output based on examples), feature-based explanation can communicate either local or global interpretation (i.e., how AI behaves specifically in a particular instance or generally on all features) depending on using which XAI techniques \citep{Molnar2020-uo}.

Local feature contribution, which is also named as feature salience, highlights input features showing their contribution for different output classes. In other words, it tells us how each feature directly affects a model to make a negative or positive prediction for the test instance \citep{Ribeiro2016-dl, Molnar2020-uo}. In contrast, global feature importance or named feature relevance, would rank input features based on their overall importance that influence the model behavior at large \citep{Lundberg2017-nj, Molnar2020-uo}. In other words, it indicates how the model relies on different features for making future predictions in general. These two types of feature-based explanations work in a complementary fashion to communicate how AI works based on feature attributions. 

\subsection{Counterfactual Explanations} 
\label{sec:CounterExpl}

Counterfactual explanations characterize a hypothetical scenario telling us what could have happened if something changes \citep{Verma2024-qb}. By adapting this scenario to describe the model's input-output relationship, counterfactual explanations help inform the smallest change to the input values that alter the model prediction \citep{Verma2024-qb}. \citet{Molnar2020-uo} highlights that the presentation of counterfactual explanations are always example-based as they have to be different instances showing how to change the test instance. 

Based on the above definition, the differences between counterfactual explanations and several types of example-based explanations described earlier could become much clearer. First, we can infer that comparative examples are always counterfactual but counterfactual explanations do not need to be actual examples existed in the training data. They can be hypothetical examples by changing feature values. More importantly, counterfactual explanations infer actionable changes, as \citet{Verma2024-qb} said ``[they] do not explicitly answer `why' the model made a prediction; instead, they provide suggestions to achieve the desired outcome.'' This suggests that the counterfactual effect of a comparative example is demonstrated by informing people what features have been changed, thereby altering model prediction. Only displaying a comparative example without highlighting the feature changes from the original example might not result in a perceivable and understandable counterfactual effect.

\subsection{Rule-based Explanations}
Rule-based explanations involve explaining model predictions by outlining one or several decision rules \citep{Molnar2020-uo}. These rules might take the form of a simple ``if...then...'' statement or more complex logic using a combination of arithmetic operators, such as ``AND...>...=...''. These rules provide a clear and structured way for humans to understand how the model arrived at a specific decision. 

Rule-based explanations can inform either local or global interpretations, depending on the specific techniques used to generate these rules \citep{Aghaeipoor2023-cd}. For example, interpretable rule-based machine learning models, like decision rules, are trained on all instances to extract rule sets as model decision-making conditions \citep{Molnar2020-uo}. In contrast, model-agnostic techniques like Anchors \citep{Ribeiro2018-ip} create rules that explain the relationship between inputs and outputs for instances close to a specific test instance. These nearby instances are created using a perturbation approach. By focusing on these neighboring instances, these rules only characterize how certain features of the test instance specifically affect the prediction, regardless of changes in the other feature values.    


\subsection{Concept-based Explanations}
\label{sec:ConceptExpl}
Concept-based explanations primarily present user-defined concepts embedded within the model's latent space to explain its predictions in the form of natural language or a ranking of concept significance \citep{Kim2018-ox, Molnar2020-uo}. For example, a zebra (the image class predicted by AI) is generally identified by its black-and-white stripes (identified concepts). 

Concept-based explanations shed light on a model's global behavior by recognizing and extracting shared semantic features for a particular class, such as colors or patterns, from the model embedding. More recently, local approaches are proposed, such as identifying the most influential regions of an image then associating these regions with broader, global concepts \citep{Shukla2023-ea} or using prompt-based image segmentation model to automatically detect concepts within an image \citep{Sun2023-vu}.  

\section{Coding Scheme for Rhetorical Strategies from Reviewed Articles} \label{appendix:codebook}

We annotated the collected design papers (\textit{N} = 41) using our rhetorical framework (Table \ref{tab:rhetorical AI explanations}), which distinguishes between two goals, explaining how AI works and explaining why AI merits use, and three rhetorical appeals: logos, ethos, and pathos. As described in Section \ref{sec:Framework}, a structured three-step was performed involving both open and axial coding. Table~\ref{tab:coding scheme} presents our coding scheme, which outlines the rhetorical elements and their illustrative mappings from the reviewed papers, along with data exemplars. Each piece of content was assigned a code indicating both the design intent it represents and the rhetorical appeal it conveys. The primary coding was conducted by the first author and subsequently validated by the second author.


\begin{longtable}{@{}>{\raggedright\arraybackslash}p{0.22\textwidth} 
                  >{\raggedright\arraybackslash}p{0.74\textwidth}@{}}
\caption{Reviewed XAI Design Papers: Mapping Design Strategies to Rhetorical Appeals}
\label{tab:coding scheme} \\
\toprule
\textbf{Paper (\textit{N}=41)} & \textbf{Mapping User Effect to Rhetorical Appeals}\\
\midrule
\endfirsthead
\multicolumn{2}{c}{{\bfseries Table \thetable\ continued from previous page}} \\
\toprule
Paper & Mapping User Effect to Rhetorical Appeals \\
\midrule
\endhead

\textbf{\citet{Alqaraawi2020-ns}} & \textbf{How AI Works: Isolated presentation} \\
 \textit{Logical} & 
Participant mentions of saliency maps features to understand the system: \\
&
p280: ``\textit{...The rationale for this was that we aimed to compare how frequently participants mentioned concepts related features that saliency maps could potentially highlight...Our results show that when saliency maps were shown, participants predicted the outcome of the classifier significantly more accurately...However, even with the presence of saliency maps, success rates were still relatively low (60.7\%)}''
p282-283: ``\textit{...Furthermore, it is hard to get a quantifiable measure of the importance of individual features in an image. Again complexity increases if one attempts to quantify the importance of a feature on new images...Another reason why noticing Saliency-features does not necessarily facilitate a better understanding of a model is that general-attributes (e.g. colour, contrast) might influence the classification outcome...}''\\
\midrule

\textbf{\citet{Alvarado2022-ys}} & \textbf{How AI Works: Isolated presentation; \newline Why AI Merits Use: Expressive interaction} \\
\textit{Credible} &
Qualitative feedback about the increased ownership of using the system: \\
&
p20: ``\textit{Center in User-Perceived Control and Awareness: ...Moreover, participants described an enhanced sense of control and even `ownership' over the recommendation system because TUIs gave them a `sense of touching, of power in their hands'.}'' \\
\textit{Emotional} &
Qualitative feedback about the enjoyability of using the system: \\
&
p17: ``\textit{During the post-interview questionnaires, P15 supported these notions: `Playing around with them [RD and RP] will very likely improve the experience with the recommendations and, therefore, with the potential enjoyability of the movies/shows.'}'' \\
\midrule

\textbf{\citet{Bansal2019-ig}} & \textbf{How AI Works: Adaptive presentation} \\
\textit{Logical} &
Effects of explanations on human mental modal of the error boundary: \\
&
p4-5: ``\textit{...in our controlled experiments we can vary the complexity of the error boundary and measure how it affects the accuracy of humans modeling the true error boundary....In the beginning, the worker makes more mistakes (more red crosses) because the mental model thus far is only partially correct. Eventually, the worker learns the correct model and successfully compensate for the AI (more red checks)...}'' p6: ``\textit{...Since error boundary complexity increases with the number of conjunctions, we observe that a more parsimonious error boundary (i.e., a single conjunction) results in a higher team performance.}'' \\
\midrule

\textbf{\citet{Bansal2019-lw}} & \textbf{How AI Works: Adaptive presentation} \\
\textit{Logical} &
Effects of explanations on human mental modal of AI predictions: \\
&
p2431: ``\textit{By analogy, we define that an update to an AI component is locally compatible with a user’s mental model if it does not introduce new errors and the user, even after the update, can safely trust the AI’s recommendations.}'' p2433: ``\textit{...Figure 3a shows that, as the number of features increases, team performance decreases because it becomes harder to create a mental model...Figure 3b shows that as errors become more stochastic, it becomes harder to create a mental model, deteriorating team performance.}''\\
\midrule

\textbf{\citet{Bansal2020-zw}} & \textbf{How AI Works: Adaptive presentation; Why AI Merits Use: Technical Assurance} \\
\textit{Logical} &
Effects of explanations on human agreement on AI predictions: \\
&
p9: ``\textit{Across the three datasets, Adaptive explanations successfully reduced the human's tendency to blindly trust the AI (i.e., decreased agreement) when it was uncertain and more likely to be incorrect.}'' \newline
Qualitative feedback: \\
&
p11(table 3): ``\textit{...Used AI as a starting references point...Double-checked after they made their own decisions...}'' \\
\textit{Credible} &
Qualitative feedback: \\
&
p11(table 3): ``\textit{...Explicitly acknowledged they used the confidence...Was more likely to accept AI above the threshold}'' \\
\midrule

\textbf{\citet{Ben_David2021-kn}} & \textbf{How AI Works: Isolated presentation, Protective framing; \newline Why AI Merits Use: Explanation as recommenders} \\ 
\textit{Logical} &
Effects of explanations on self-reported readiness to adopt and willingness to pay: \\
&
p394-5: ``\textit{On the very first day when the model was introduced (day 3) we see an average of 60\% adoption when averaging over all participants...Following the model failure on day 7, adoption plummeted, returning approximately to the initial level with 61\% on day 8, this was followed by a slow recovery up until day 10 (66\%)...Finally, when participants were told advice will no longer be free, and were asked about willingness to pay, we see an overall decrease in adoption on the final day (77\%).}'' \\
\textit{Emotional} &
Algorithmic aversion: \\
&
p397: ``\textit{...the two more elaborate explanations, the Performance-Based (P-value = 0.0445, t = -2.0225) and the Feature-Based (P-value =0.0498, t = -1.9741) prove to be more resistant to the effect of algorithm error...When we explore the adoption recovery...we find that the Performance-Based explanation had a marginally significant (P-value = 0.0633, t = 1.8691) tendency to recover better than the global explanation (12\% compared to only 2\% for the Feature-Based and the Global Explanation, respectively)...}''\\
\midrule

\textbf{\citet{Bernardo2023-xv}} & \textbf{How AI Works: Isolated presentation, Explanation affect, Emotional cues} \\
\textit{Logical} &
Effects of explanations on self-reported usefulness: \\
&
p15: ``\textit{Example-based explanation increases interestingly surprised, trusting emotions, and perceived usefulness, while feature-and rule-based explanation decreases them. Logic robotic increases fearfully dismayed plus anxiously suspicious emotions and decreases perceived usefulness, while humanized communication functions the opposite. The presence of supplementary information decreases fearfully dismayed emotions, while the absence of it increases the effect.}'' \\
\textit{Credible} &
Effects of explanations on self-reported trust and reliance (see above summary findings) \\
\textit{Emotional} &
Participant AI anxiety and incidental emotion influence user perceived trust of explanations: \\
&
p16: ``\textit{First, when fearfully dismayed emotions were felt, users with high AI anxiety experienced lower perceived trust than those who have low AI anxiety}.''
\\
\midrule

\textbf{\citet{Bove2022-pd}} & \textbf{How AI Works: Mechanistic presentation} \\
\textit{Logical} &
Effects of explanations on self-reported system understanding: \\
&
p814: ``\textit{We propose a questionnaire approach...Each item in the questionnaire is a statement, for which users can either answer "true", "false" or "I don’t know". We design three types of questions to capture different components of user understanding: (i) explanations' scope...(ii) explanations' effects...(iii) explanations' locality}'' p816: ``\textit{Contextualization improves objective understanding...Exploration does not have a significant impact...The interaction of contextualization and exploration does not have a significant impact neither.}''
\\
\textit{Credible} &
Effects of explanations on self-reported satisfaction: \\
&
p816: ``\textit{Contextualization significantly improves users satisfaction...contextualization parameter is greater than the claimed value and conclude that contextualization significantly improves non-expert users’ satisfaction...the exploration parameter is greater than the claimed value and conclude that exploration significantly improves non-expert users’ satisfaction}''
\\
\midrule

\textbf{\citet{Cai2019-lb}} & \textbf{How AI Works: Isolated presentation} \\
\textit{Logical} &
Effects of explanations on self-reported system understanding, capability, and benevolence: \newline
p3-4``\textit{When the algorithm failed to recognize the drawing, users who received normative explanations felt they had a better understanding of the system...Likewise, when the algorithm did not recognize the drawing, users also rated system capability higher if they had seen a normative explanation...In addition, we found a small but significant main effect of comparative explanations on system benevolence. Users who saw comparative explanations perceived the system to be more benevolent.}'' \\
\midrule

\textbf{\citet{Cai2019-zx}} & \textbf{How AI Works: Mechanistic presentation; \newline Why AI Merits Use: Explanation as retriever} \\
\textit{Logical} &
Effects of explainable interactive tool on self-reported diagnostic utility, future use, and overall preference: \newline
P8(figure 5): ``\textit{Compared to a conventional interface, SMILY had greater diagnostic utility and required lower effort. Users also indicated having greater trust in SMILY, felt SMILY offered better mental support, and felt they were more likely to use it in clinical practice.}'' \\
&
Qualitative feedback: \newline
p10-11: ``\textit{Tracking the Likelihood of a Decision Hypothesis. During the study, users described ways in which iteratively refining search results helped them track the likelihood of a hypothesis...Generating New Ideas. The refinement tools also provided decision-making support by helping users reflect on their thought process and generate new ideas...Reducing Problem Complexity. Reference images containing a mixture of features were often cognitively difficult to analyze...}'' \\
\textit{Credible} &
Effects of explanations on self-reported trust in the system: \\
&
p9: ``\textit{In sum, users were able to retrieve more diagnostically relevant information with less effort using SMILY, had greater trust in the system, and felt they were more likely to use it in clinical practice...}''
\\
\midrule

\textbf{\citet{Cau2025-rj}} & \textbf{How AI Works: Mechanistic presentation; Why AI Merits Use: Technical assurance} \\
\textit{Logical} &
Effects of explanations on accuracy and overreliance: \newline
p1449: ``\textit{...Consequently, we fail to reject the null hypotheses for H1a and H1b, concluding that the explanation type does not affect users’ accuracy and overreliance on AI...}''\\
&
Self-reported understanding and usefulness of explanations: \\
&
p1450-1451: ``\textit{...15 participants perceived Local Model Explanations (LME) as fast and time-saving support for accepting/rejecting an application...12 participants highlighted the difficulty in understanding why the AI assigned specific weights to attributes and the heavy emphasis on the computer skills attribute...10 participants were overwhelmed by excessive information and suggested improving the visualization of the data distribution module.}'' \newline
Qualitative feedback: \\
&
p1451: ``\textit{...10 participants were overwhelmed by excessive information and suggested improving the visualization of the data distribution module...}''
\\
\textit{Credible} &
Self-reported user confidence and trustworthiness of explanations to use AI: \\
&
p1453: ``\textit{...our exploratory analysis revealed an association between requesting on-demand assistance and lower user confidence. This suggests that accessing data or AI support can undermine confidence, regardless of the type of assistance provided. However, it remains unclear whether explanations reduced confidence or if participants requested assistance due to feeling less confident.}'' \newline
Qualitative feedback: \newline
p1451: ``\textit{...The Data-Centric Explanations (DCE) were perceived by 18 participants as trustworthy, giving clear guidelines for the job application requirements...}''\\
\midrule

\textbf{\citet{Chen2023-vc}} & \textbf{How AI Works: Isolated presentation} \\
\textit{Logical} & 
Effect of explanations on decision-making accuracy and reliance: \newline
p15: ``\textit{...Across tasks, example-based explanations achieved complementary decision performance whereas feature-based explanations did not...}''
Qualitative feedback: \newline
P13: ``\textit{With feature-based explanations, participants applied intuition about features to determine if they agreed or disagreed with the model’s reasoning...example-based explanations can provide additional context and support inductive reasoning}'' \newline
Combined summary: \newline
``\textit{Example-based explanations led to less disruption of people’s natural intuition about the outcome...Example-based explanations better supported recognizing or even learning new features because the similar examples and their ground truth labels brought in additional context that could promote inductive reasoning...Example-based explanations provided strong signals of unreliability, particularly when the AI system predicts both examples incorrectly.}'' \\
\midrule

\textbf{\citet{Diaz-Rodriguez2020-es}} & \textbf{Why AI Merits Use: \newline Authority framing, Cultural relatedness} \\
\textit{Credible} &
Qualitative feedback: \newline
p319: ``\textit{...Does embedding of expert/domain knowledge into DL models help explain such models? Can XAI help encode such prior knowledge? Juan Jesus Pleguezuelos, History teacher and podcast author of Art History for entrance exams to University: The challenge I pose is to make others see an historical image only through words. It is clear that this requires an exhaustive description of the masterpiece, but you should also try to make others feel the latent soul in it, and decipher the intention of the author...}''\\
\textit{Emotional} &
Qualitative feedback: \newline
p319: ``\textit{...And if you could also convey the emotion that this work is able to cause, it can be that words may be more than enough to make a listener understand an artistic work that he is not seeing in that moment...}''\\
\midrule

\textbf{\citet{Ebermann2023-hx}} & \textbf{How AI Works: Isolated presentation; \newline Why AI Merits Use: Protective framing} \\
\textit{Logical} &
Self-reported evaluation of AI support: \newline
p1815: ``\textit{This suggests that cognitive misfit reduces the subjective support. However, in contrast to our expectations, the evaluation improved from T2 to T3 in the case of decision situations with cognitive misfit...value on the variable `evaluation of AI's support' was higher at the third time point of measure...The `evaluation of AI's support' is indeed lower when cognitive misfit occurs.}''
\\
\textit{Credible} &
Self-reported willingness to use AI: \newline
p1816: ``\textit{In the case of cognitive misfit resulting from different decisions and corresponding explanations of the AI and the user, users do not have significantly less willingness to use the AI again in comparison to cases where cognitive fit occurs.}''
\\
\textit{Emotional} & 
Self-reported mood score: \newline
p1815: ``\textit{In line with our expectation...where the decision and explanation of the AI is known and cognitive dissonance might occur, that the values of the variable `mood' are significantly lower in the case of participants in decision situations with cognitive misfit...}''
\\
\midrule

\textbf{\citet{Ehsan2021-id}} & \textbf{How AI Works: Isolated presentation; Why AI Merits Use: Social endorsement} \\
\textit{Credible} & 
Qualitative themes: \newline
``\textit{Supported Effects...Understanding organizational norms and values, social validation, transitive trust, and temporal relevance...}''\\
\midrule

\textbf{\citet{Ehsan2024-lu}} & \textbf{How AI Works: Isolated presentation, Numerical association} \\
\textit{Logical} &
Self-reported intelligence and understandability: \newline
``\textit{Participants with AI background also viewed numbers as potentially actionable even when their meaning was unclear...Unable to access their meaning, the non-AI group associated numbers with the presence of a higher, more intelligent expression...both AI and non-AI groups found unanticipated explanatory value in AD’s declarative statements and NR’s numerical representations...}''
\\
\textit{Credible} &
Self-reported confidence, friendliness, and second chance use: \newline
``\textit{Participants in both groups had unwarranted faith in numbers. However, their extent and reasons for doing so were different...The mere presence of numbers was associated with an algorithmic thinking process in the robot even when the meaning of numbers was unclear...The AI group seems to have followed heuristic reasoning that associates mathematical representations with logic and intelligence...}''
\\
\midrule

\textbf{\citet{Evans2022-zg}} & \textbf{How AI Works: Isolated presentation, Graphical synergy; \newline Why AI Merits: Explanation as recommenders} \\
\textit{Logical} &
Self-report on ``explanations are understandable'' and ``explanations are useful for decision-making'': \newline
p286: ``\textit{Explanation examples are presented in descending order of total positive responses (slightly agree to strongly agree)to the statement: `The explanation provides me with valuable information for my work.' Trust scores were the most widely accepted explanation by this metric, while the explanation receiving the most positive median response was the Counterfactuals (One-axis) example.}'' \newline
Qualitative feedback: \newline
P288-289: ``\textit{...Two saliency map examples difficult to understand, distracting and/or confusing (p1, 2, 5), to trivially understandable but not at all useful (p6)...The risk for positive confirmation bias with such [saliency] explanations was pointed out: ``[It] is hiding a lot of information. But it's very tempting, because it’s very simple'' (p1)}" \newline
P289: ``\textit{These [concept] examples were widely considered intuitively understandable and informative about the factors important for the AI output, albeit with limited value ascribed to this information (p1, 2, 4, 6)...Criticisms of the explanation were the lack of precision and/or descriptiveness (p1–3), that having to read text is generally undesirable in the pathologist workflow (p3, 4)}'' \newline
P290: ``\textit{Pathologists were almost unanimous in expressing that these [counterfactuals] explanations `[helped them] understand what the algorithm is looking for' (p1) All participants drew some concrete conclusions about the factors important to the AI solution, finding it self-explanatory that staining was the most important factor distinguishing positive nuclei from those marked negative, with some identifying other important factors such as shape, size of nuclei, particularly in separating the positively annotated from the unclassified nuclei (in the two-axis example) (p1–3).}'' \\
\textit{Credible} &
Qualitative feedback: \newline
p290: ``\textit{The labeling of annotations as high- or low-confidence was generally regarded as being intuitively understandable. Aside from identifying the AI solution’s confidence in its annotations, many participants also inferred from this the factors that may have been important to the solution.}''\\
\midrule

\textbf{\citet{Freire2023-gh}} & \textbf{How AI Works: Isolated presentation, Emotional cues} \\
\textit{Credible} &
Effects of emoji and explanations on human-AI agreement: \newline
``\textit{The explainable AI (XAI)-emoji and XAI conditions have higher agreement scores than the emoji and control (no intervention) conditions.}'' \\
\midrule

\textbf{\citet{Gaole2025-wo}} & \textbf{How AI Works: Adaptive and Mechanistic presentation} \\
\textit{Logical} &
Effects of explanations on user objective feature understanding and appropriate reliance based on human-AI agreement:\\
&
p914-915: ``\textit{We found thatbLLM Agent condition achieved significantly worse objective feature understanding than the Dashboard, CXAI, and ECXAI conditions...In our study, we found that interactive XAI interfaces can have a negative impact of increasing over-reliance on the AI system.}''
\\
\textit{Credible} &
Effects of explanations on self-reported trust of the AI system and user confidence: \\
&
p915: ``\textit{With post-hoc Tukey's HSD test, we found that participants who received XAI showed significantly higher trust in Understandability/Predictability (i.e., Control < Dashboard, CXAI, ECXAI)}.''
\\
\textit{Emotional} &
Effects of explanations on self-reported engagement of using the system (\textit{No direct report in the result})
\\
\midrule

\textbf{\citet{Kenny2021-xb}} & \textbf{How AI Works: Isolated presentation; \newline Why AI Merits Use: Explanation as retriever} \\
\textit{Logical} & 
Effects of explanation on correctness: \\
&
p17: ``\textit{...explanations seem to mainly have an effect when things go wrong, when errors arise and outputs diverge from what a user expects/desires...}'' \newline
Self-reported reasonableness of explanations: \\
&
p21: ``\textit{...this system-level measure of correctness (like the trust measure) shows that the item-level effects for Explanation do not persist into a system-level evaluation of correctness. The Error-Rate's variable was also found to impact overall correctness and overall reasonableness; see Table 2), showing people are sensitive to changes in error-rates from 4\% to 12\% and 28\%.}''
\\
\textit{Credible} &
Self-reported trust and satisfaction: \\
&
p21: ``\textit{...Specifically, the Explanation-Present group rated Question 1 (“From the explanation, I understand how the program works”), Question 3 (“This explanation of how the program works has sufficient detail”), and Question 5 (“This explanation of how the program works tells me how to use it”) significantly higher than the Explanation-Absent group, all p < .05...}''\\
\midrule

\textbf{\citet{Kim2024-hp}} & \textbf{How AI Works: Mechanistic presentation; \newline Why AI Merits Use: Explanation as recommenders} \\
\textit{Logical} &
Qualitative feedback: \\
&
p7-8: ``\textit{...clinicians found the interfaces helpful in making informed decisions and determining the order of importance of  different aspects...To enhance their experience, patients suggested some improvements, including the provision of visual cues to facilitate a more intuitive understanding of their current status. One patient mentioned the following regarding the color-coding scheme...The clinicians expressed a strong desire to identify the types and causes of errors made by their patients...}''\\
\midrule

\textbf{\citet{Kocielnik2019-sq}} & \textbf{How AI Works: Isolated presentation; \newline Why AI Merits Use: Expectation management} \\
\textit{Logical} &
Self-reported understanding of AI model, perceived control over system behaviors:\\
&
p8: ``\textit{Comparing conditions with and without Explanation revealed a significant positive impact...with higher average level of perceived understanding (mean of both understanding questions) for participants who saw the Explanation than for those who did not...We therefore consider H2.3 supported as the Control Slider intervention significantly increased participants' perceived level of control over the Scheduling Assistant’s behavior.}'' \\
\textit{Emotional} &
Expected satisfaction of system performance: \\
&
p10: \textit{Participants in the High Recall version of the system reported no significant difference in satisfaction between the Baseline and Pure-techniques. At the same time, in the High Precision version, participants exposed to Pure-techniques reported significantly higher satisfaction than those in the Baseline...} \\
\midrule

\textbf{\citet{Liu2021-ds}} & \textbf{How AI Works: Isolated presentation with interactivity}\\
\textit{Logical} &
Effects of explanations on appropriate trust (reliance): \\
&
p22: ``\textit{In general, we do not find significant impact from interactive explanations with respect to the performance of human-AI team or human agreement with wrong AI predictions, compared to static explanations. However, humans are more likely to find AI assistance useful with interactive explanations than static explanations in ICPSR and COMPAS, but not in BIOS...}'' \newline
Self-reported real-time assistance useful: \\
&
p25: \textit{In COMPAS, Interactive/Interactive achieves a significantly higher human perception of real-time assistance usefulness than both Static/Static and Interactive/Static...Perception of Interactive/Static is also significantly higher than that of Static/Static. We find similar results in ICPSR except that the difference between Static/Static and Interactive/Static is not significant...} \newline
Self-reported identification of important features: \\
&
p25: ``\textit{These results suggest that participants with interaction are more likely to fixate on demographic features and potentially reinforce human biases,11 but are less likely to identify computationally important features in ICPSR and COMPAS.}''
\\
\midrule

\textbf{\citet{Ma2023-rj}} & \textbf{How AI Works: Adaptive presentation; Why AI Merits Use: Technical assurance} \\
\textit{Logical} &
Effects of explanation on human-AI agreement (appropriate trust): \\
&
``\textit{Figure 6 shows the human-AI agreement in different human-AI CL situations under three conditions. Results showed that all three conditions made people’s agreement with AI significantly higher when AI’s CL was higher than that when the human’s CL was higher...the human-AI agreement in AI Confidence was significantly higher than Adaptive Workflow , and Adaptive Recommendation, and it was marginally higher than Direct Display. Note that when AI is wrong, a lower agreement with AI is better. These results suggest humans’ less over-trust in AI in our proposed CL exploitation conditions...However, when the AI’s recommendation was correct, we did not observe significant differences in human-AI agreement between the three proposed conditions and AI Confidence baseline.}'' \newline
Qualitative feedback: \\
&
``\textit{Participants in Adaptive Workflow and Adaptive Recommendation were forced to think independently.}'' \\
\textit{Credible} &
Qualitative feedback of correctness likelihood: \\
&
``\textit{Some participants doubted the displayed CL information and ignored it in the decision-making process...Most participants referred to or were influenced by the dis- played CL or CL-based adaptation}'' \\
\midrule

\textbf{\citet{Nguyen2018-az}} & \textbf{How AI Works: Isolated presentation with interactivity} \\
\textit{Logical} &
Effect of presenting AI predictions on fact-checking accuracy: \newline
p194: ``\textit{We observe that there are larger changes for errors made by participants in group Control: those who have not seen our stance predictions change their answers more than those who have...This roughly corresponds to the errors by the veracity classifier, showing again that system predictions can both help users (when correct) or lead users to errors that reflect model fallibility or biases implicit in training data.}'' \newline
Qualitative feedback: \newline
p196: ``\textit{It was very hard to understand. It seems on one task, I was 100\% sure it was true and I was told it was false, I even followed links to verify the sources}''\\
\textit{Emotional} &
Qualitative feedback: \newline
p196: ``\textit{...thought it was really cool! I’d enjoy playing with this more if it wasn’t during my work time...}'' \\
\midrule

\textbf{\citet{Okolo2024-et}} & \textbf{How AI Works: Isolated presentation, Cultural relateness, Emotional cues}\\
\textit{Logical} &
Qualitative feedback: \\
&
p11: ``\textit{While these interpretations made sense to the CHWs, they were not what the XAI methods intended to convey (they intended to show the original image for reference and the annotated image for explainability, with no storyline among the images). Only a few CHWs could partially understand the explanations with the help of higher-level features such as colors, images, or shapes}''
p17: ``\textit{This interpretation was incorrect because the underlying model relied on specific bodily regions (in alignment with the Kramer’s Rule chart in Table 1) to provide a severity prediction, not on the number of yellow boxes.}''
\\
\textit{Credible} &
Qualitative feedback: \\
&
p13: ``\textit{P04 and P05 emphasized that their patients trust them more when they use visual aids. They gave the example of paper-based visual aids they carry to educate pregnant women about when to seek urgent care. When we asked if CHWs would prefer similar paper-based visual aids for jaundice, P04 and P05 preferred the AI-driven app over paper-based visual aids, stating that the app could diagnose the severity of diseases, which the paper-based aids could not.}''
\\
\textit{Emotional} &
Qualitative feedback: \\
&
p8: ``\textit{They were worried that the red color might confuse the participants since it is often used in public health messaging to imply `danger' in the medical sense.}''
p14: ``\textit{CHWs often relied on their prior experience with colors in visual aids in the domain of Public Health to interpret what the colors in the XAI visualizations might mean. They had strong color associations, believing that `green means safe, yellow means warning, red means danger and gray/blue means lack of blood.' These color associations largely influenced how they interpreted the visualizations depicting a jaundiced baby.}''
\\
\midrule

\textbf{\citet{Okoso2025-ja}} & \textbf{How AI Works: Authority framing, Emotional cues} \\
\textit{Credible} &
Self-reported acceptance of AI advice: \\
&
``\textit{In the Formal, Authoritative, and Humorous groups, an upward trend was observed in d tone u, indicating that participants tended to increase their scores. Although no significant overall effect of tone was observed, the tone could possibly affect participants differently depending on specific user attributes.}'' \newline
Qualitative feedback: \\
&
``\textit{Among the 135 participants who noticed a change in tone and provided relevant responses, 59 participants said it did not affect their decisions, whereas 55 participants said it did. Some participants felt that an informal tone was inappropriate for a serious topic, such as recidivism risk, whereas others found the tone easier to understand.}'' \\
\textit{Emotional} &
Qualitative feedback: \\
&
``\textit{A total of 110 participants (37\%) expressed either positive or negative opinions about the explanations but ultimately stated that it did not influence their decision. Most felt that the explanations did not provide sufficient information relevant to their needs...those movies based solely on genre and did not consider the explanation, whereas others remarked that it felt unnatural for an AI, which supposedly lacks emotions, to promote a movie}''\\
\midrule

\textbf{\citet{Schemmer2023-zy}} & \textbf{How AI Works: Isolated presentation} \\
\textit{Logical} &
Effects of explanation on appropriate reliance: \\
&
p418: ``\textit{In the XAI condition, we can observe a significant increase in relative AI reliance (PAIR) from 29.59\% to 38.87\% while the (relative self-reliance) RSR does not change significantly (71.87\%) for the control condition and 69.45\% for the feature importance condition. This means explanations of AI decisions can reduce the share of under-reliance. It is important to highlight that PAIR is not increased simply by relying more often on AI advice, as this would have also reduced the RSR significantly.}''
\\
\midrule

\textbf{\citet{Schuff2022-ly}} & \textbf{How AI Works: Isolated presentation} \\
\textit{Logical} &
Self-reported importance of features reflect bias of understanding explanations: \\
&
p619: ``\textit{Overall, three studies confirmed the presumably biasing influence of word length, (pairs of) two studies respectively confirmed the effect of sentence length, display index and saliency rank, and one study (each) found significant effects of word position, sentiment polarity, word frequency, capitalization and dependency relation}''
\\
\midrule

\textbf{\citet{Sun2023-dm}} & \textbf{How AI Works: Isolated presentation; Why AI Merits: Explanation as retriever}\\
\textit{Logical} &
Improve model performance by using local explanation-based attention steering and evaluate the model vulnerability: \\
&
p19-20: ``\textit{S2 results suggest that (1) the workflow of the local explanation-based attention steering provided a diverse perspective in diagnosing model vulnerability, (2) the direct steering design helped the process of model revision straightforward, and (3) every participant enjoyed improved key model performance measures...The behavioral data we collected shows that all participants generated the model that outperforms (1) its model accuracy, (2) the overlap between the model’s focus and the relevant object types (IoU), and (3) the proportion of reasonable attention out of all images in a test set...After completing the user studies, the majority of users strongly agree that adjusting local explanations can effectively improve model performance.}'' \newline
Qualitative feedback: \\
&
p20: ``\textit{During interviews, all participants shared their positive impressions about the effectiveness of attention adjustment in improving model accuracy, which is the primary objective of conducting model fine-tuning. They also confirmed that the impact of contextual bias was reduced as attention quality increased by attention steering}'' p22: ``\textit{Existing issues: insufficient description about reasonability matrix, Ill-formed local explanation visualization...}''
\\
\midrule

\textbf{\citet{Suresh2022-lv}} & \textbf{How AI Works: Isolated presentation with interactivity; \newline Why AI Merits Use: Explanation as retriever}\\
\textit{Logical} &
Effects of explanation on human-AI agreement: \\
&
p776: \textit{We recorded the percent of cases in which participants agreed with the model (versus when they disagreed or were not sure). For cases in which the prediction was correct, the agreement rate was similar across conditions; however, when the prediction was incorrect, we found that participants were less likely to accept the model’s prediction when they were using the NN interface, with or without the input editor. Often in these cases, they did not explicitly “disagree” with the model, but wanted additional information about the signal and/or patient before committing to an answer.} \newline
Qualitative feedback: \\
&
p776: ``\textit{Nearest neighbors enable reasoning with clinically-relevant concepts...In contrast, with the baseline condition, participants often had difficulty extracting higher-level, clinically-relevant concepts from the feature importance visualization...Nearest neighbors help characterize uncertainty and incorporate it into decision-making}'' \\
\midrule

\textbf{\citet{Van_der_Waa2021-he}} & \textbf{How AI Works: Isolated presentation; Why AI Merits Use: Explanation as recommender} \\
\textit{Logical} &
Identification of model decisive factor and advice: \\
&
p15: ``\textit{First, rule-based explanations indeed seem to allow par- ticipants to more accurately identify the factor from a situation that was decisive in the system’s advice. However, rule-based nor example-based explanation allowed participants to learn to predict system behavior.}'' \newline
Effects of explanations on perceived system understanding and prediction of correctness: \\
&
p15: ``\textit{The rule-based explanations however, did cause to participants to think that they better understood the system compared to example-based and no explanations. The example-based explanations only showed a small and insignificant increase in perceived system understanding.}'' \\
\midrule

\textbf{\citet{Vasconcelos2023-cm}} & \textbf{How AI Works: Isolated presentation}\\
\textit{Logical} &
Effects of explanations on overreliance and the influence of task difficulty: \newline
p14: ``\textit{an increase in task difficulty increases the cost of following the strategy to engage with the task, as it generally takes more cognitive effort for people to either (a) ensure that the AI has gotten the answer correct when provided with only predictions or (b) complete the task alone. However, the strategy to rely on the AI does not incur the same costs, as people are not engaging with how hard the task is.}'' p21: ``\textit{the salient explanation condition has less overreliance than all other conditions (prediction, highlight explanation, written explanation, and incomplete explanation...We also find that the incomplete explanation condition has less overreliance than the prediction only and written explanation conditions, more overreliance than the salient explanation condition and finds no difference in overreliance in the highlight explanation condition.)}.''\\
\midrule

\textbf{\citet{Wambsganss2020-fq}} & \textbf{How AI Works: Adaptive and Mechanistic presentation; \newline Why AI Merits Use: Explanation as recommender} \\
\textit{Logical} &
Self-reported intention to use, perceived usefulness and ease of use: \\
&
p9: ``\textit{the mean value of intention to use of participants using AL as a writing tool was 2.33 (SD= 0.58). These values are significantly better than the results of the alternative scripting approach. For perceived usefulness we observed a mean value of 3.2 (SD= 1.12) and for perceived ease of use the value was 2.83 (SD= 1.08) for participants from the control group. The mean value for the intention to use was 3.5 (SD= 1.13). The results clearly show that the participants of our experiment rated the acceptance of AL as an adaptive feedback tool positively compared to the usage of the alternative application. The statistical significance was also proven in a double-sided t-test for all three constructs (see table 5).}'' \newline
Qualitative feedback: \\
&
p9: ``\textit{The general attitude for AL was very positive.  Especially the fast and direct feedback, the graph-like visualization  of the argumentation structure and the summarizing scores  were mentioned quite often. However, sometimes AL was not  correctly classifying claims and premises, which users suggest to improve.}''\\
\midrule

\textbf{\citet{Wang2019-we}} & \textbf{How AI Works: Mechanistic presentation; 
\newline Why AI Merits Use: Explanation as recommender}\\
\textit{Logical} & 
Qualitative feedback: \\
&
p10-11: ``\textit{Finding Alternative Hypotheses. We found many instances of users seeking alternative hypotheses with or without various types of XAI...P4 found counterfactual reasoning useful even though one cannot change a patient’s history, 'because it’s a matter of possibility. It’s how  you rank your differentials...}'' \\
\textit{Credible} &
Qualitative feedback: \\
&
p11: ``\textit{Lack of Trust: Verify with Supplementary, Situational, Source Data. A lack of trust drove users to want to verify the AI’s decision by looking at raw data, in this case, patient case and physical exam notes, or recorded data...without the context of the patient, [she] can’t diagnose based on the vitals itself. So [she] wouldn’t really trust}'' \\
\midrule

\textbf{\citet{Wang2021-hv}} & \textbf{How AI Works: Isolated presentation} \\
\textit{Logical} &
Effects of explanations on human understanding of model behaviors and calibrated trust: \\
&
p325-326: ``\textit{The only positive effect of model explanation that we have consistently observed across different decision making tasks is that feature importance explanations increase people’s objective understanding of an AI model, while feature contribution explanations increase people’s subjective understanding of an AI model...feature contribution explanation seems to be able to satisfy more desiderata of AI explanations when people have some domain expertise in the decision making task...For the two example-based explanations in our study, we found minimal evidence on their capability to support trust calibration.}'' \\
\midrule

\textbf{\citet{Wang2023-qb}} & \textbf{How AI Works: Adaptive presentation} \\
\textit{Logical} &
Effects of explanations and model updates on perceived system understanding: \\
&
p8: ``\textit{We find that participants’ perceived change of explanations increased as the the explanations of the updated model in Phase 2 became more dissimilar from those of the old model used in Phase 1. In other words, participants in our experiment could perceive the change in model explanations brought up by a model update.}'' \newline
Effects of explanations and model updates on objective just (frequency of accepting model decisions): \\
&
p8: ``\textit{participants’ objective trust gain and subjective trust gain in the AI model from Phase 1 to Phase 2, both across all participants and within subgroups of participants with different levels of understanding of the AI model. However, we find that neither participants’ objective trust nor their subjective trust seems to be affected by the similarity level of model explanations between Phase 1 and Phase 2.}''
\\ 
\textit{Credible} & 
Effects of explanations and model updates on subjective trust gain and satisfaction gain: \\
&
See above excerpt about objective trust; p8: ``\textit{Figure 3(d) further shows participants’ subjective satisfaction gain from Phase 1 to Phase 2, conditioned on their understanding score. Still, participants did not seem to significantly change their satisfaction with the AI model as the similarity of model explanations before and after the update varied.}''\\
\midrule

\textbf{\citet{Wang2023-ar}} & \textbf{How AI Works: Mechanistic presentation, Emotional cues; Why AI Merits Use: Explanation as recommender}\\
\textit{Logical} &
Effects of counterfactual explanations on several tool useful measures: `easy to use,' `easy to understand,' `help understand model,' `help find ways to improve plans,' and `will apply loan again': \\
&
p11: ``\textit{(A) Participants thought GAM Coach was relatively easy and enjoyable to use, and the tool helped them identify actions to obtain a preferred ML decision. (B) All interaction techniques, especially experimenting with hypothetical values, were rated favorably.}'' \newline
Qualitative feedback: \\
&
p9: \textit{Three main reasons that participants reported choosing plans were that the plans were (1) controllable, (2) requiring small changes or less compromise, or (3) beneficial for life in general...Participants’ explanations for not choosing a plan mostly complemented the reasons for choosing a plan...Preference configuration is helpful...Participants’ explanations for not choosing a plan mostly complemented the reasons for choosing a plan. Some participants also skipped plans because they were puzzled by counterintuitive suggestions, did not understand the suggestions, or just wanted to see more alternatives.}\\
\textit{Emotional} &
Effects of explanations on enjoyable to use (\textit{See previous excerpt on tool usefulness in logical})\\
\midrule

\textbf{\citet{Yang2020-wm}} & \textbf{How AI Works: Isolated presentation}\\
\textit{Logical} &
Effects of explanation and spatial layout on user appropriate trust of AI predictions: \\
&
p195: ``\textit{The results support that all the visual explanations largely increase appropriate trust, reduce overtrust, and help users gain more confidence in their decisions...image-based explanations help correct undertrust, but the effect of rose-based explanations on undertrust is inconclusive...For the image-based explanations, all the three spatial layouts lead to a similar level of users’ trust; appropriate trust, overtrust, and undertrust are very similar and display few differences among the three spatial layouts}'' \newline
Effects of explanation on perceived helpfulness: \\
&
\textit{Helpfulness was measured but not directly reported in results} 
\\
\textit{Credible} &
Effects of explanation on self-confidence, trust meter before and after AI feedback: \\
&
p196-197: ``\textit{Before seeing the feed- back, participants increase the trust meter for a correct recommendation and decrease the trust meter for an incorrect one; having an explanation shows small positive effects on trust direction...Image-based explanations outperform rose-based explanations because they increase appropriate trust, decrease overtrust and undertrust, improve self-confidence, and show more usability}''
\\
\midrule

\textbf{\citet{Zhang2020-pf}} & \textbf{How AI Works: Isolated presentation} \\
\textit{Logical} &
Effects of explanation on human-AI agreement: \\
&
``\textit{The baseline condition (green) and the explanation condition (blue) had similar agreement percentages, while the with-confidence condition (orange) had higher percentage when the confidence level was above 70\%.}''
\\

\bottomrule
\end{longtable}

\end{document}